\documentclass[twocolumn] {book}
\usepackage{times}
\usepackage{graphicx}
\usepackage{lscape}
\usepackage{xcolor}
\usepackage{amsmath}
\usepackage{amssymb}
\usepackage{calrsfs}
\usepackage{bm}
\usepackage{natbib}
\usepackage[margin=1.0in,bottom=0.8in,top=0.8in]{geometry}
\renewcommand{\cite}{\citet}

\begin{document}

\label{firstpage}

\setcounter{chapter}{8} 
\chapter[]{\bf Spatial And Temporal Changes Of The Geomagnetic Field: Insights From Forward And Inverse Core Field Models}

Nicolas Gillet\footnote{Univ. Grenoble Alpes, Univ. Savoie Mont Blanc, CNRS, IRD, IFSTTAR, ISTerre, 38000 Grenoble, France}

\subsection*{Abstract}

Observational constraints on geomagnetic field changes from interannual to millenial periods are reviewed, and the current resolution of field models (covering archeological to satellite eras) is discussed. 
With the perspective of data assimilation, emphasis is put on uncertainties entaching Gauss coefficients, and on the statistical properties of ground-based records. 
These latter potentially call for leaving behind the notion of geomagnetic jerks. 
The accuracy at which we recover interannual changes also requires considering with caution the apparent periodicity seen in the secular acceleration from satellite data. 
I then address the interpretation of recorded magnetic fluctuations in terms of core dynamics, highlighting the need for models that allow (or pre-suppose) a magnetic energy orders of magnitudes larger than the kinetic energy at large length-scales, a target for future numerical simulations of the geodynamo. 
I finally recall the first attempts at implementing geomagnetic data assimilation algorithms. 

\subsection*{keywords}

geomagnetic secular variation, geodynamo simulations, inverse problem, data assimilation, Earth's core dynamics

\section[]{Introduction}

The past two decades have seen our knwoledge on the geomagnetic field and the dynamics within the Earth's core strongly modified by the new possibilities offered by modern computers, and the inflow of continuous satellite observations, in complement of ground-based records. 
We are able to run geodynamo simulations that ressemble to the Earth: some are dipole dominated, and under some conditions they show polarity reversals \citep{jones2015thermal}. 
At the same time, as we push the parameters towards Earth-like values, numerical geodynamos do not show reversals anymore \citep{schaeffer2017turbulent}.  
We indeed face several severe obstacles: no simulation is currently able to mimic all time-scales together, and 
diffusive processes remain too large. 
We also suffer from limited geophysical constraints.  
Direct geomagnetic records \citep[20 yr of continuous satellite data, less than 200 yr of observatory series, four centuries of historical measurements, see][]{jackson2015geomagnetic} cover a short era in comparison with key time-scales of the core dynamics, while we know from paleomagnetic studies that changes occur at all periods up to some 100 kyr \citep{constable05}. 
The available window is also cut towards high frequencies, since periods shorter than $O(1)$ yr are hidden by signals from external (ionospheric and magnetospheric) sources \citep[e.g.][]{finlay2017challenges}. 

In this chapter I review the main recent findings concerning time changes of the geomagnetic field on periods ranging from a couple of years to millenia, covering both observational constraints, dynamical modeling, and the combination of the two through data assimilation studies. 
It is on purpose that I exclude mechanical forcings (precession, tides...). 
These are undoubtedly a source of core motions, and an altenative scenario to planetary dynamos \citep{le2015flows}. 
However, up to now they failed at producing Earth-like dynamos.  
This may just be a question of time, as much more efforts have been put on the convective side.   

I first recall the main dynamical features captured with geomagnetic data: from interannual to decadal changes with modern direct records, to centennial and longer fluctuations as seen with historical data, archeological artefacts or sediment series (\S\ref{sec: long periods}-\ref{sec: SA}). 
I highlight methodological issues associated with the complex separation of magnetic sources, and the difficult estimation of model uncertainties (\S\ref{sec: separation}-\ref{sec: uncertainties}).  
Then I present some milestones in terms of time-scale separation if one wishes to reach Earth-like dynamics, and some current limitations of geodynamo simulations (\S\ref{sec: time-scales}). 
Several reduced models are next illustrated (\S\ref{sec: MC}-\ref{sec: MAC}), developed on purpose to analyse core field changes from interannual to centennial periods, based on Magneto-Coriolis (MC) or Magneto-Archimedes-Coriolis (MAC) waves.
In \S\ref{sec: dynamos} I recall important findings from numerical simulations that integrate primitive equations. 
I report in \S\ref{sec: assimilation} efforts carried out to merge the information contained into geophysical observations with that coming from dynamical models, a process known as data assimilation: 
attempts at using geodynamo simulations for either static or dynamical recoveries of the core state (\S\ref{sec: snapshot}-\ref{sec: assim dynamo}), and applications to the construction of field models (\S\ref{sec: assim applications}).
I finally recall crucial issues in \S\ref{sec: conclusion}. 

\section[]{Geomagnetic Field Modeling}
\label{sec: observations}

\subsection{Separation Of Sources}
\label{sec: separation}

We shall ignore electrical currents in the mantle \citep[a condition valid for periods longer than a few months, see][]{jault:2015}, 
so that the magnetic field ${\bf B}$ derives from a potential: ${\bf B}=-\nabla V$. 
From the solenoidal condition $\nabla\cdot{\bf B}=0$, $V$ satisfies to Laplace's equation, $\nabla^2V=0$. 
In absence of electrical currents, internal and external solutions arise, 
which in spherical coordinates ($r,\theta,\phi$) are naturally expressed by means of spherical harmonics \citep{olsen2010separation} and Gauss coefficients (denoted by $g_n^m$ and $h_n^m$, with $n$ and $m$ the degree and order). 

A proper separation of internal and external signals requires the knowledge of several components of the field on a closed surface. 
This is unfortunaltely never completely achieved. 
Ignoring some components of the field leads to non-uniqueness issues \citep{sabaka2010mathematical}.
The coverage is obviously un-even for archeomagnetic, observatory and historical records, in favor of lands (resp. seas) for ground-based stations \citep[resp. ship logs, see][]{jonkers03}. 
Such disparities in the spatial sampling imply ambiguities between Gauss coefficients \citep[see][for an application to the observatory era]{gillet13}. 
The major North/South discrepancy in the coverage by archeological artifacts and lavas translates into sensitivity kernels much biased towards the North and the Middle-East \citep[e.g.][]{korte11}. 

Magnetic records from space are available continuously from low-orbiting satellite data since 1999, with the Oersted, SAC-C, CHAMP and Swarm missions \citep{finlay2017challenges}. 
With their quasi polar trajectory, they cover an almost entire spherical surface, but the remaining polar gaps nevertheless induce larger uncertainties on zonal coefficients \citep{olsen2012}. 
Furthermore, the sampling of external signals towards low frequencies (longer than a few months) is hindered by the slow drift of the satellite orbit in local time. 
The associated difficulty in describing slow evolutions of external sources is a major limit to the recovery of rapid  (shorter than a couple of years) internal fiels changes.
Another important issue when considering satellite data is the ionosphere being internal to the spacecraft trajectory:
one thus cannot properly separate core and ionopsheric field changes by solving Laplace's equation.
Finally, the current-free separation leading to magnetic potential fields breaks down in places (such as auroral regions) where the satellite trajectory crosses electrical currents \citep{finlay2017challenges}. 

Separating the signature of the several sources in satellite and ground-based stations is carried out by means of field modeling (i.e. space-time interpollation). 
Complete descriptions often rely on subtle weighting of the many parameters entering this inverse problem \citep[e.g. the comprehensive approach by][]{sabaka2015cm5}, although attempts at objectifying the prior information now appear \citep{holschneider2016correlation}.
The absence of global deterministic physical models able to isolate the external activity also leads to consider several magnetic indices \citep[e.g.][]{olsen2014chaos,thomson2007improved}, used to select quiet time data, and onto which rapid external variations may be anchored \citep[see][]{lesur2010second,finlay2016recent}. 
  
\subsection{Field Model Uncertainties}
\label{sec: uncertainties}

Solutions for internal model coefficients are obtained by minimizing a cost function, sum of a measure of the misfit to magnetic data and a measure of the model complexity. 
In the case of quadratic norms, the penalty function takes the form
\begin{eqnarray}
J({\bf x})=\left\|{\bf y}-{\sf H}({\bf x})\right\|^2_{\sf R}
+\left\|{\bf x}-\overline{\bf x}\right\|^2_{\sf P}\,,
\label{eq: cost fn}
\end{eqnarray}
with the notation $\left\|{\bf x}\right\|^2_{\sf M}={\bf x}^T{\sf M}^{-1}{\bf x}$.
Vectors ${\bf x}$ and ${\bf y}$ respectively store model parameters and observations. 
$\overline{\bf x}$ is a background model, ${\sf H}$ is the observation operator, and matrices ${\sf P}$ and ${\sf R}$ store the a priori cross-covariances of respectively the model anomaly to $\overline{\bf x}$ and the observation error. 

Models of the main field (MF) originating from the core shall be used as input data in re-analyses of the core dynamics (see \S\ref{sec: assimilation}), a reason why it is necessary to estimate uncertainties on Gauss coefficients. 
With quadratic measures as in equation (\ref{eq: cost fn}), posterior uncertainties may be described by the inverse of the Hessian matrix (that measures the curvature of $J$ in the neighborhood of the solution ${\bf x}^*$), 
\begin{eqnarray}
{\sf C} = \left(\nabla{\sf H}({\bf x}^*)^T{\sf R}^{-1}\nabla{\sf H}({\bf x}^*)+{\sf P}^{-1}\right)^{-1}\,.
\label{eq: hessien}
\end{eqnarray}
Estimating a posteriori model uncertainties thus requires reasonable estimates of both the observation errors and the prior information on the unknown parameters (through respectively  ${\sf R}$ and ${\sf P}$). 
Unfortunately, none of these is easy to handle. 

A first concern is the numerical possibility of calculating ${\sf C}$ in cases where millions of data and/or parameters are to be considered. 
If the model size is too large \citep[e.g. when co-estimating rapidly changing external sources,][]{sabaka2015cm5}, storing ${\sf C}$ will not be possible. 
In the case of a large number of data, considering cross-covariant data errors will be out of reach, possibly leading to biases in the resulting field model. 
Alternatively, one may consider data-set reduced at given frequencies, for instance with observatory series \citep{macmillan2013observatory}, or spacially averaged as done with virtual observatories \citep{mandea2006new}.

In any case, an a priori model norm is often penalized in the inversion process, in order to ensure the spectral convergence of the model in space and time, and avoid generating too many short length/time-scale oscillations. 
This is achieved by applying regularizations, which for instance damp the second or third time derivative, on the top of a spatial norm \citep[following][]{bloxham1992time}. 
By doing so one a priori assumes that short wave-lengths are tiny, which is not based on any physical insight. 
As a consequence the a posteriori uncertainties are severely under-estimated at short length/time-scales, and the obtained model should be considered as a time-weighted estimate towards high harmonic degrees.
This drawback is partly alleviated by the use of more realistic second-order statistics when constructing the prior matrix ${\sf P}$ \citep{gillet13}, or correlation based modeling \citep{holschneider2016correlation}. 
In such cases it becomes meaningful to consider the formal posterior covariance matrix (\ref{eq: hessien}) as a measure of model uncertainties, if data errors are decently estimated. 

However, assessing observation errors also is a tricky issue. 
One should consider both measurement errors (the accuracy of the instruments) and errors associated with unmodelled sources. 
Estimating the former is not straight-forward: see the protocoles for providing standard deviations in observatory series \citep{lesur2017estimating} or uncertainties on data from the Swarm mission \citep{toffner2016flight}.  
But more importantly, the signature of unmodelled processes often dominate the error budget. 
For instance, even the most up-to-date satellite field models do not capture entirely the ionospheric processes, in particular in the auroral region \citep{finlay2017challenges}.
Another example concerns archeomagnetic records that are entached by many processes, posterior to the age of the target magnetization, and which may alter the minerals within the sample, or the orientation of the recorded field \citep[e.g.][]{constableTOG15}.
In most cases, such unmodelled sources result in biases and/or errors correlated in space and/or time. 
This prevents from considering ${\sf C}$ as a perfect measure of a posteriori model uncertainties. 
For instance, by increasing the amount $N$ of data, formal posterior errors will decrease as $1/\sqrt{N}$ (consequence of the central limit theorem). 
Will result illusory small uncertainties if such records are entached by biases. 
Alternatively, Monte-Carlo (ensemble) strategies to sample observations may be used to estimate model errors, an approach widely used in archeomagnetic field modeling \citep{panovska2015limitations}, but which does not remove the drawbacks induced by regularizations, as illustrated below. 

\subsection{Centennial To Millennial Geomagnetic Changes}
\label{sec: long periods}

\begin{figure}[t]
\centerline{
	\includegraphics[width=1\linewidth]{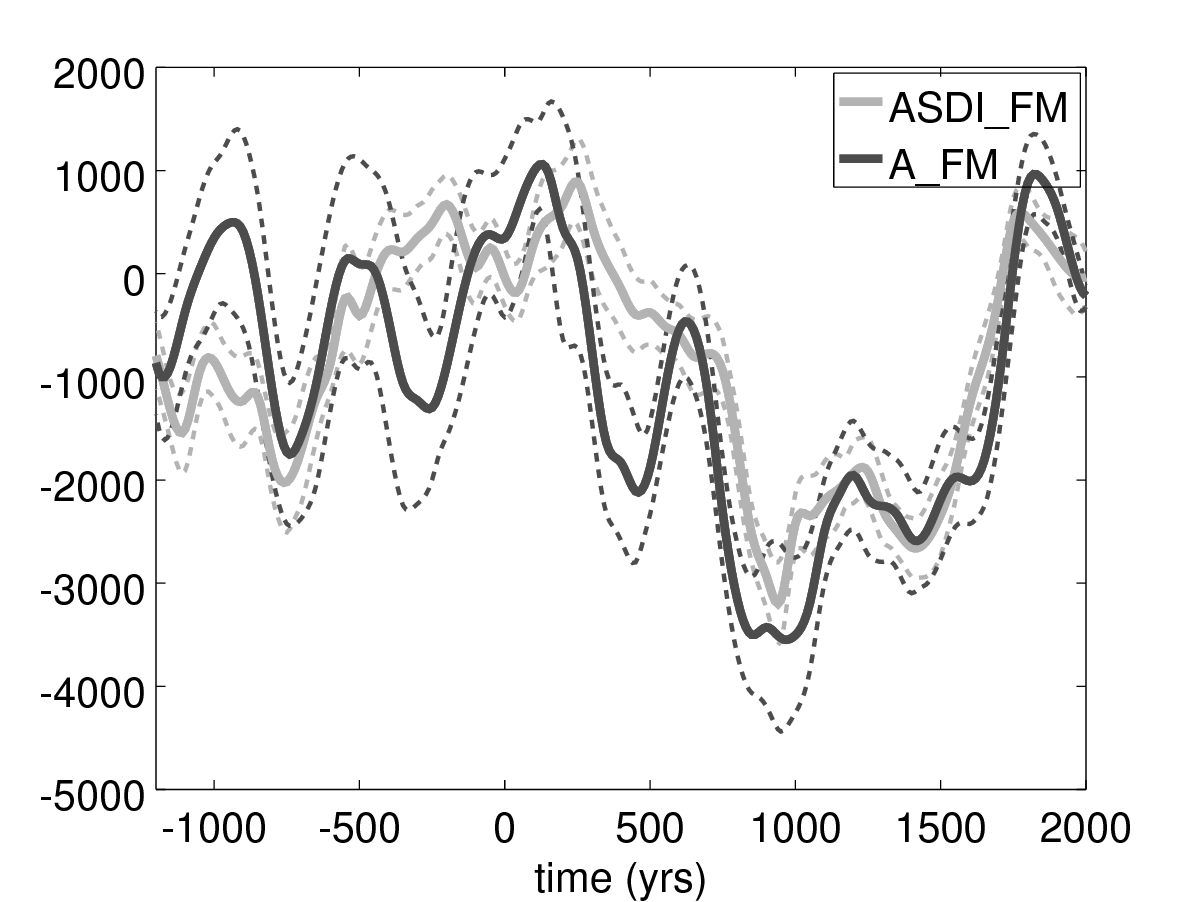}}
\centerline{
	\includegraphics[width=1\linewidth]{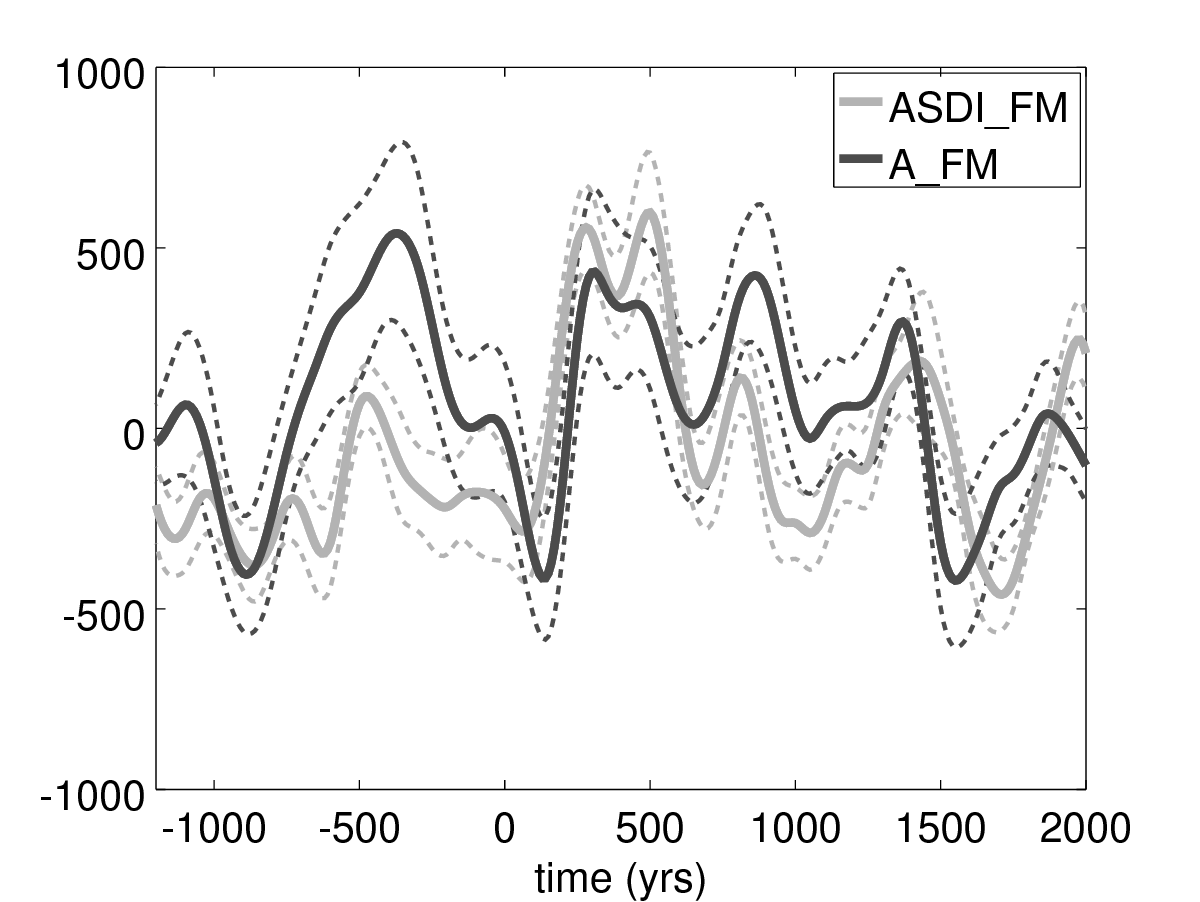}}
	\caption{Time series of Gauss coefficients $h_2^2$ (top) and $h_4^2$ (bottom), for the field models A\_FM (black) and ASDI\_FM (grey) by \cite{licht13}.
The former is built from archeological and lava data only, the latter incorporates sediment data. 
Dotted lines represent $\pm1\sigma$ uncertainties. 
}
	\label{fig: archeo oscillations}
\end{figure}

Despite these numerous difficulties (non-exhaustive, even not speaking about dating uncertainties associated with archeomagnetic artifacts and sediment cores), interesting features are seen in Gauss coefficient series of archeomagnetic field models deduced from global databases \citep[e.g. GEOMAGIA,][]{brown2015geomagia50.1}.
Such models bring out some debated, intruiguing non-dipolar features. 
Oscillations of period $\approx 250$ yrs are observed for instance in Western Europe intensity records \citep{genevey2016new}.
These suggest either complex combinations of Gauss coefficient series, or oscillations at twice this period in some specific coefficients. 
Interestingly, even though global model predictions partly filter out centennial changes seen in regional records, fluctuations of period $\approx 500$ yrs are isolated in Gauss coefficient series (see Figure \ref{fig: archeo oscillations}).
These are particularly clear for models inverted from archeological and lava records, while considering sediment series make some of the centennial fluctuation disappear (see $h_2^2$ prior to 500AD). 
According to the uncertainties provided with the field models, such oscillations are resolved.
The discrepancy between models containing or not sediment data thus calls for some explanations. 

\begin{figure}[t]
\centerline{
	\includegraphics[width=1\linewidth]{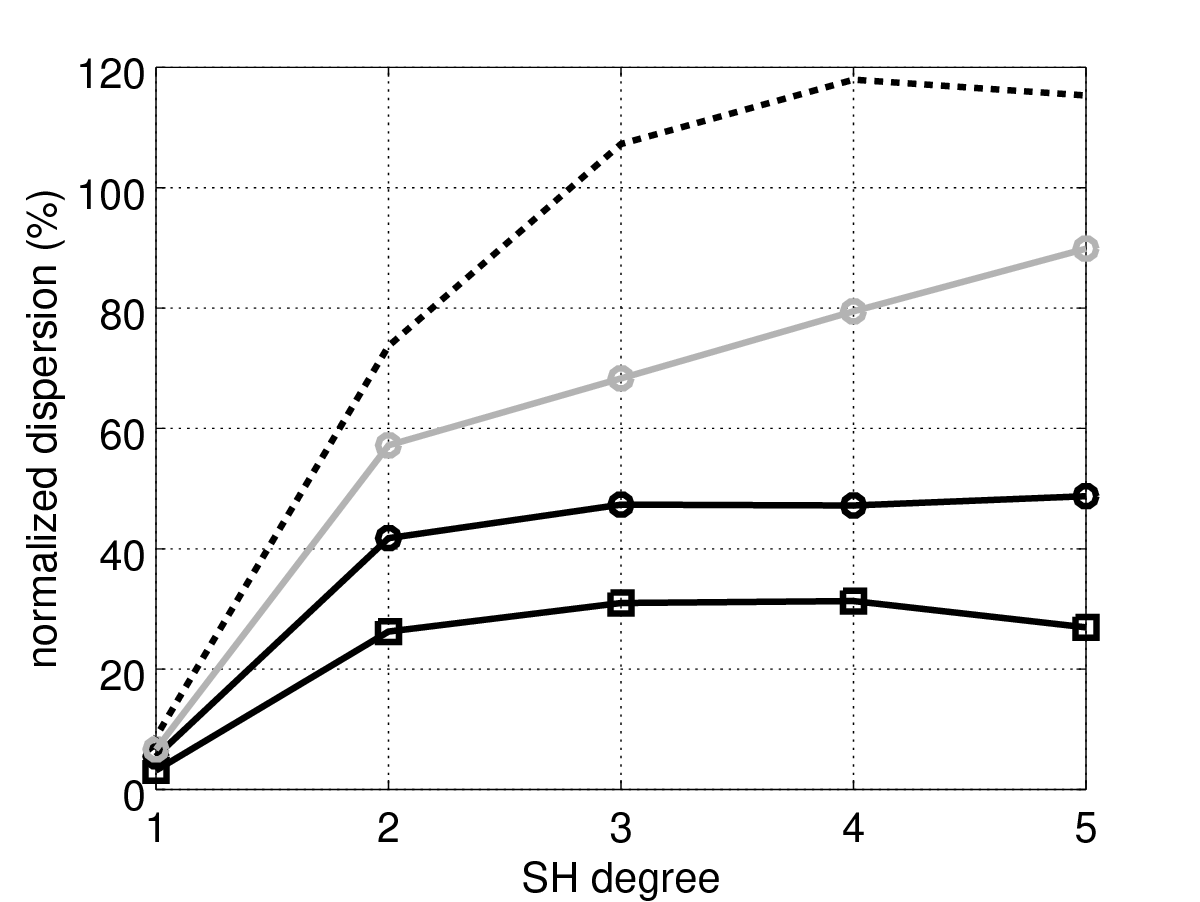}}
	\caption{Normalized dispersion $\chi(n)$, as measured with equation (\ref{eq: relative error archeo}), for the models A\_FM  and ASDI\_FM \citep[resp. black circles and squares,][]{licht13} and the model AmR \citep[grey circles,][]{sanchez2016modelling}.
The ensemble results from either bootstrap (A\_FM and ASDI\_FM) or ensemble Kalman filter (AmR) methods. 
The black dotted line shows the norm, per degree $n$, of the difference between A\_FM  and ASDI\_FM, normalized by the norm of the models. }
	\label{fig: archeo dispersion}
\end{figure}

Let consider the normalized, time-average model uncertainty per degree $n$, as measured by
\begin{multline}
\chi(n)=\left[%\frac
\left({\displaystyle\int_{t_s}^{t_e}\sum_{i=1}^{N_e}\sum_{m=0}^n {g_n^m}_i(t)^2+{h_n^m}_i(t)^2\mathrm{d}t}\right)^{-1}
\right.\\\left.
{\displaystyle\int_{t_s}^{t_e}\sum_{i=1}^{N_e}\sum_{m=0}^n \left( {g_n^m}_i(t)-\left<g_n^m\right>\right)^2+\left( {h_n^m}_i(t)-\left<h_n^m\right>\right)^2\mathrm{d}t}
\right]^{1/2},
\label{eq: relative error archeo}
\end{multline}
for an ensemble of $N_e=20$ realizations. 
Brackets denote the ensemble average. 
We logically witness (Figure \ref{fig: archeo dispersion}) a decrease in $\chi$ when incorporating sediment cores on the top of archeological data.
However, $\chi(n)$ saturates towards high degrees, consequence of the damping used in the model construction (see \S\ref{sec: uncertainties}): uncertainties tend to be under-estimated towards small length-scales. 
Furthermore, the model uncertainty is lower than the difference between models built with and without sediment data: in at least one of these two models, posterior errorbars must be too low. 
Alternatives to this issue may be found (see \S\ref{sec: assim dynamo}), with the use of stochastic priors \citep{hellio18} or of dynamo norms \citep{sanchez2016modelling}. 
These latter spatial constraints provide larger posterior uncertainties in comparison with regularized models, as shown in Figure \ref{fig: archeo dispersion}, reaching about 100\% error at degree 5. 

But the prior information implicitely introduced through the damping is not the only responsible for the too large discrepancy between models incorporating or not sediment data. 
As highlighted by \cite{licht13}, assessing the relative weigths of archeological and sediment data is not straightforward.
The sedimentation rate associated with each core actually tends to smooth in time the magnetic signal recorded in sediments \citep{pavon2014geomagnetic}. 
Added to the large dating errors, this process acts as a low-pass filter. 
As a consequence, sediment records should be considered as time-weighted data. 
Accounting for such a process in the construction of the field models is feasible as far as algorithms are concerned. 
However, this would involve estimating, for each sediment core, a time-dependent filter translating sedimentation rates into time-weighting functions. 
 
Sediment and archeological data also indicate signals on longer periods, although the existence of clearly isolated periodicities is debated \citep{nilsson2011millennial}. 
Given the short era of observatory data, historical databases \citep{jonkers03,arneitz2017histmag} are the only remaining source to study centennial field changes to a higher accuracy. 
The most strinking time-dependent pattern, extracted from the {\it gufm1} field model \citep{jackson00}, is the westward drift of intense flux patches in the equatorial belt. 
Mainly symmetric with respect to the equator, and the most obvious at azimuthal wave number $m=5$, they travel at a speed of about 17 km/yr, which translates into a period around 275 yrs \citep{finlay03}.
Physical mechanisms candidates to explaining such features are presented in \S\ref{sec: dynamics}.  

\subsection{Getting Rid Of Geomagnetic Jerks?}
\label{sec: jerks}
 
The era of modern magnetic measurements (observatory and satellites) has brought several surprises. 
Short magnetic series may appear boring at first sight. 
However, what actually matters, as far as its main source (the dynamics within the core) is concerned, is actually the secular variation $\partial {\bf B}/\partial t$ (SV). 
Indeed, this latter is directly related to the fluid velocity ${\bf u}$ within the core, through the induction equation
\begin{eqnarray}
\frac{\partial {\bf B}}{\partial t}=\nabla\times\left({\bf u}\times{\bf B}\right)+\eta\nabla^2{\bf B}\,,
\label{eq: induction}
\end{eqnarray}
where $\eta\sim 1$ m$^2$/s is the magnetic diffusivity of the metallic core.
For this reason, SV series are scrutinized, which put the focus on the so-called geomagnetic jerks \citep[see][for a review]{mandea2010geomagnetic}, or sudden changes in the rate of change of the field. 

\begin{figure*}[t]
\centerline{
	\includegraphics[width=0.5\linewidth]{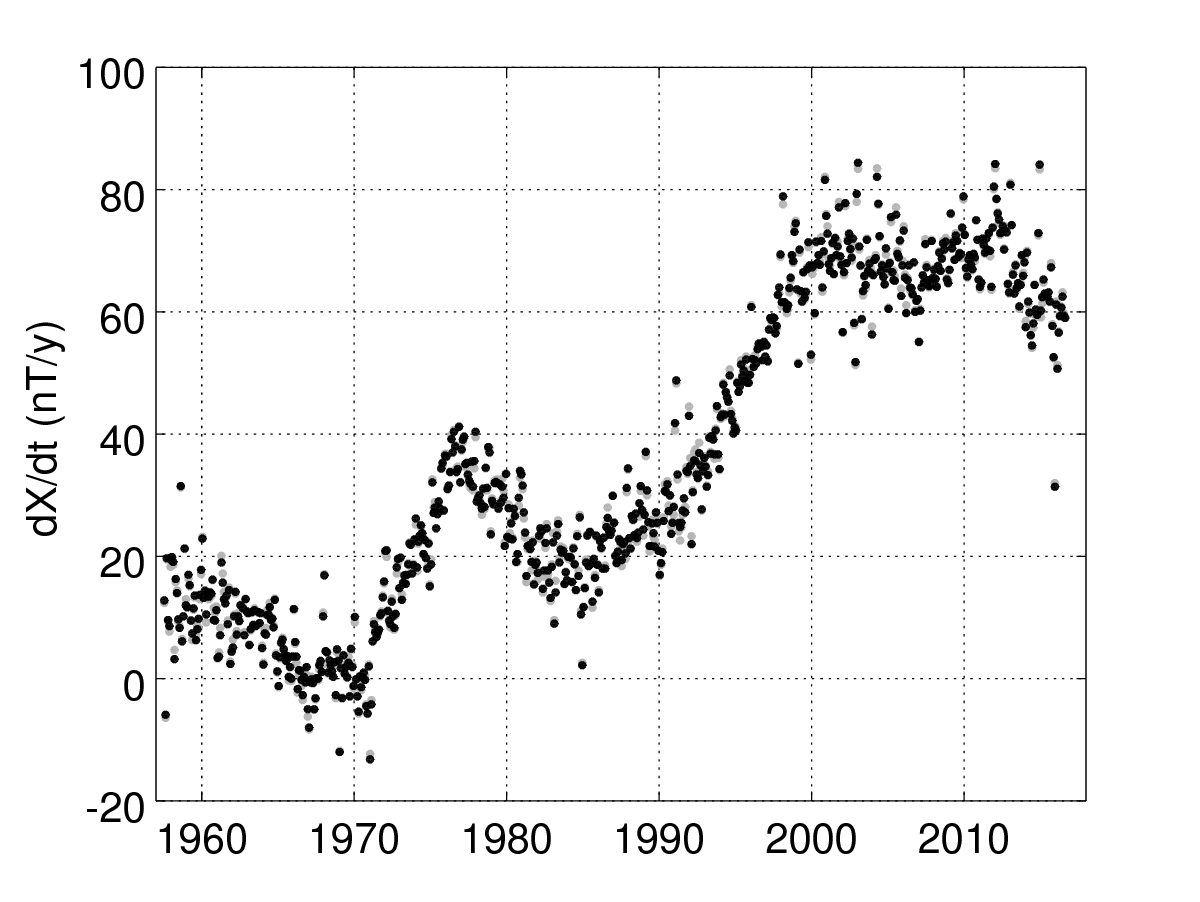}
	\includegraphics[width=0.5\linewidth]{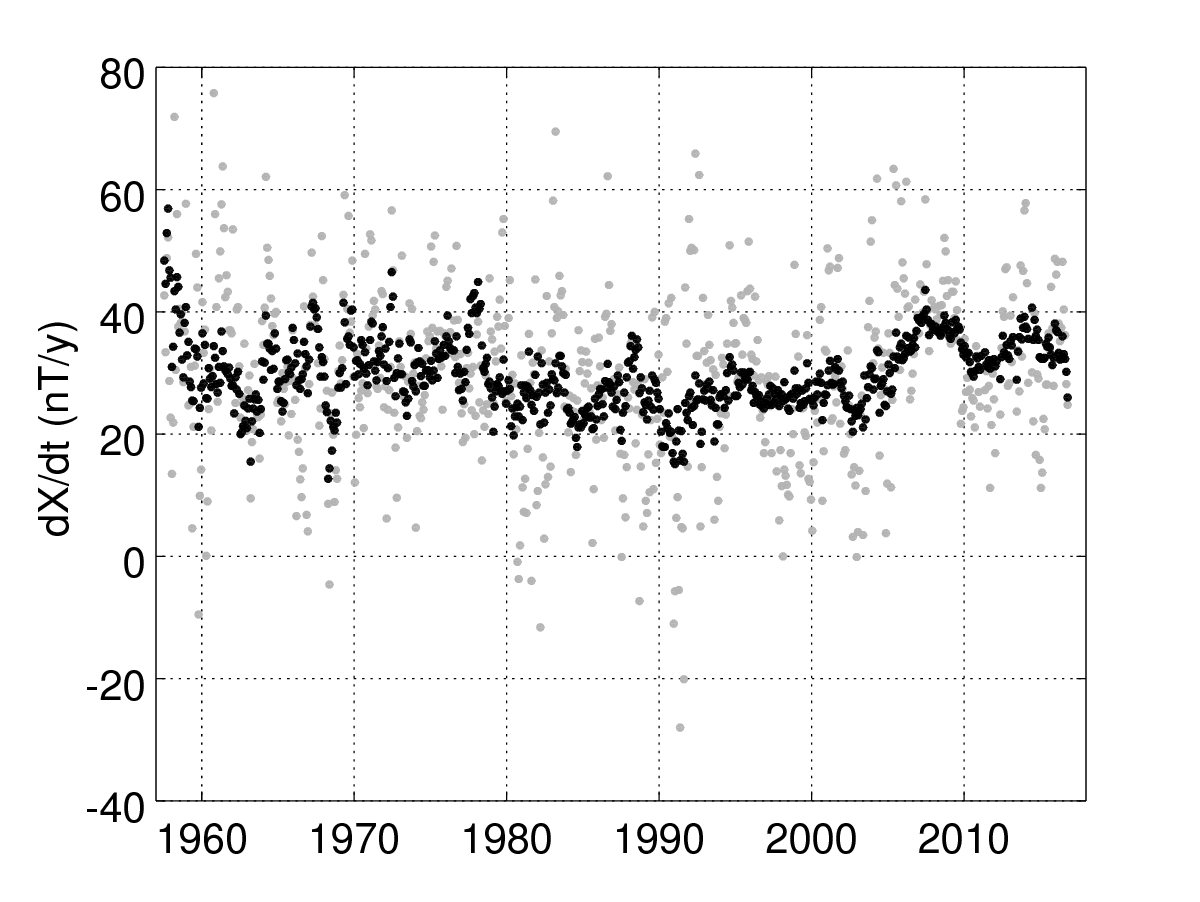}
}
\centerline{
	\includegraphics[width=0.5\linewidth]{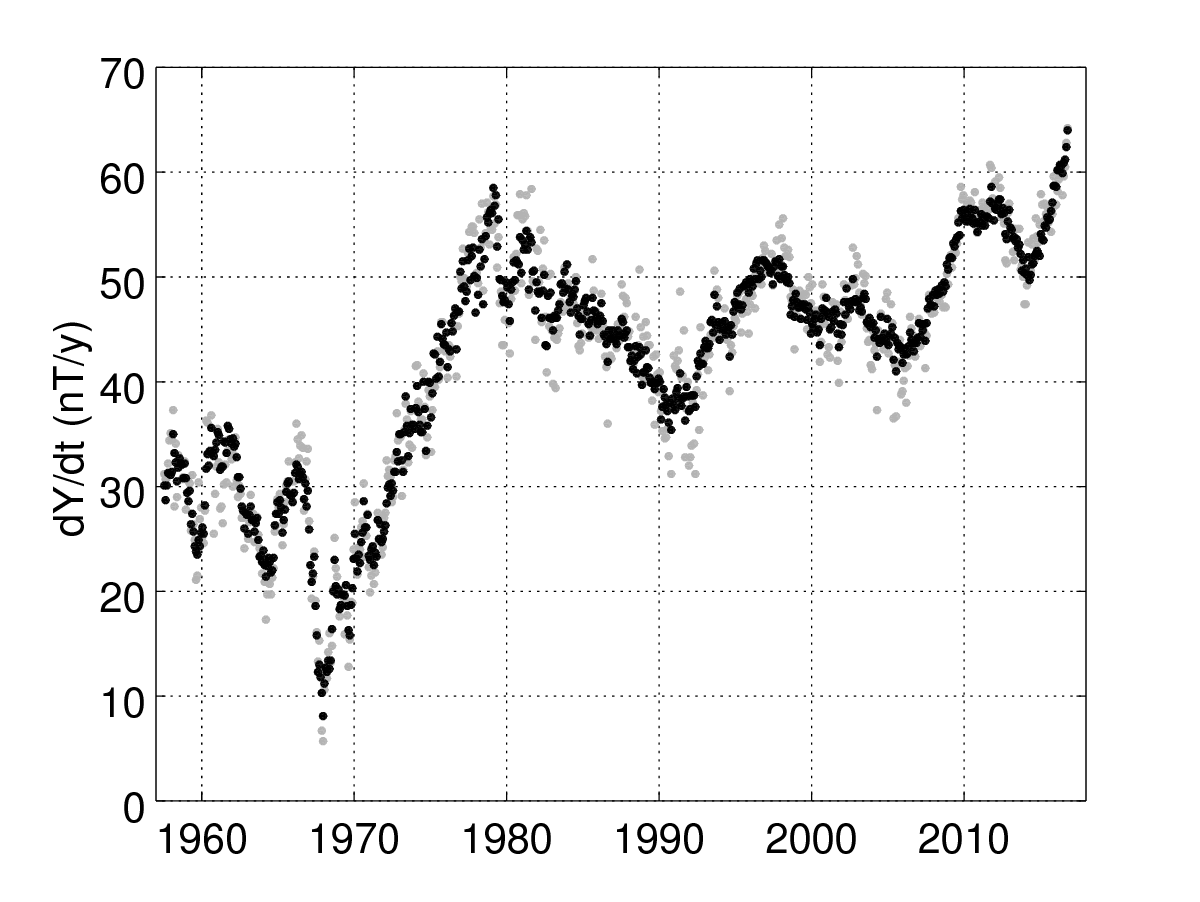}
	\includegraphics[width=0.5\linewidth]{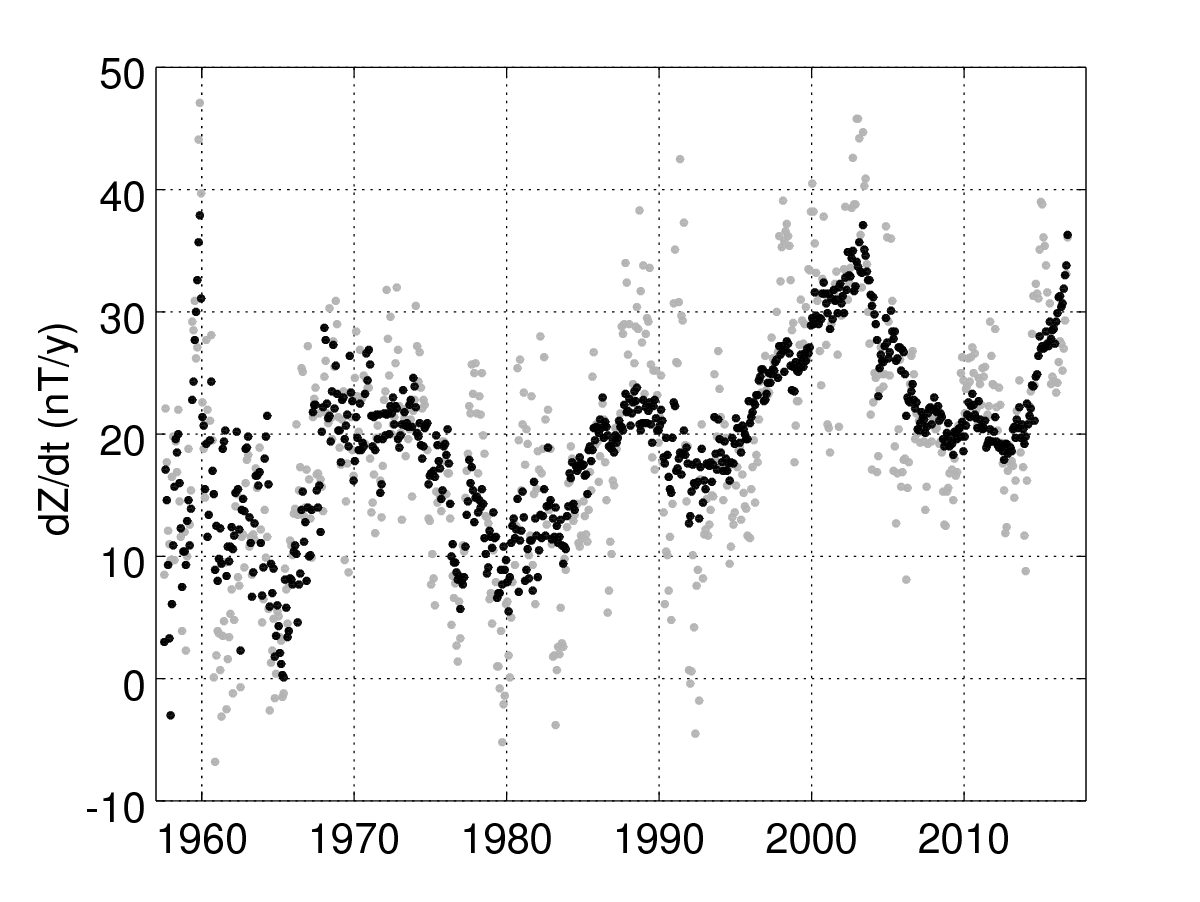}
}
	\caption{SV series (annual differences of monthly means) from ground-based data. 
Top: $dX/dt$ at Thule ($77^{\circ}$N, $69^{\circ}$W, left) and M'Bour ($14^{\circ}$N, $17^{\circ}$W, right). 
Bottom: $dY/dt$ (left) and $dZ/dt$ (right) at Chambon-la-For\^et ($48^{\circ}$N, $2^{\circ}$E). 
$(X,Y,Z)$ are local cartesian coordinates, respectively Northward, Eastward and downward. 
In grey ordinary monthly values calculated from hourly means. 
In black the revised monthly means \citep{olsen2014chaos}: CHAOS magnetospheric field model predictions (using as input a preliminary extended version of the RC index), and CM4 ionospheric field predictions \citep[][using as input the F10.7 index]{sabaka2004extending}, are first removed from hourly means, before a Huber-weighted robust monthly mean is computed.
Hourly mean data (ftp://ftp.nerc-murchison.ac.uk/geomag/Swarm/AUX\_OBS/hour/) have been produced by the British Geological Survey for ESA within the framework of the Swarm mission \citep{macmillan2013observatory}. 
}
	\label{fig: SV changes}
\end{figure*}

\begin{figure}[t]
\centering
	\includegraphics[width=1\linewidth]{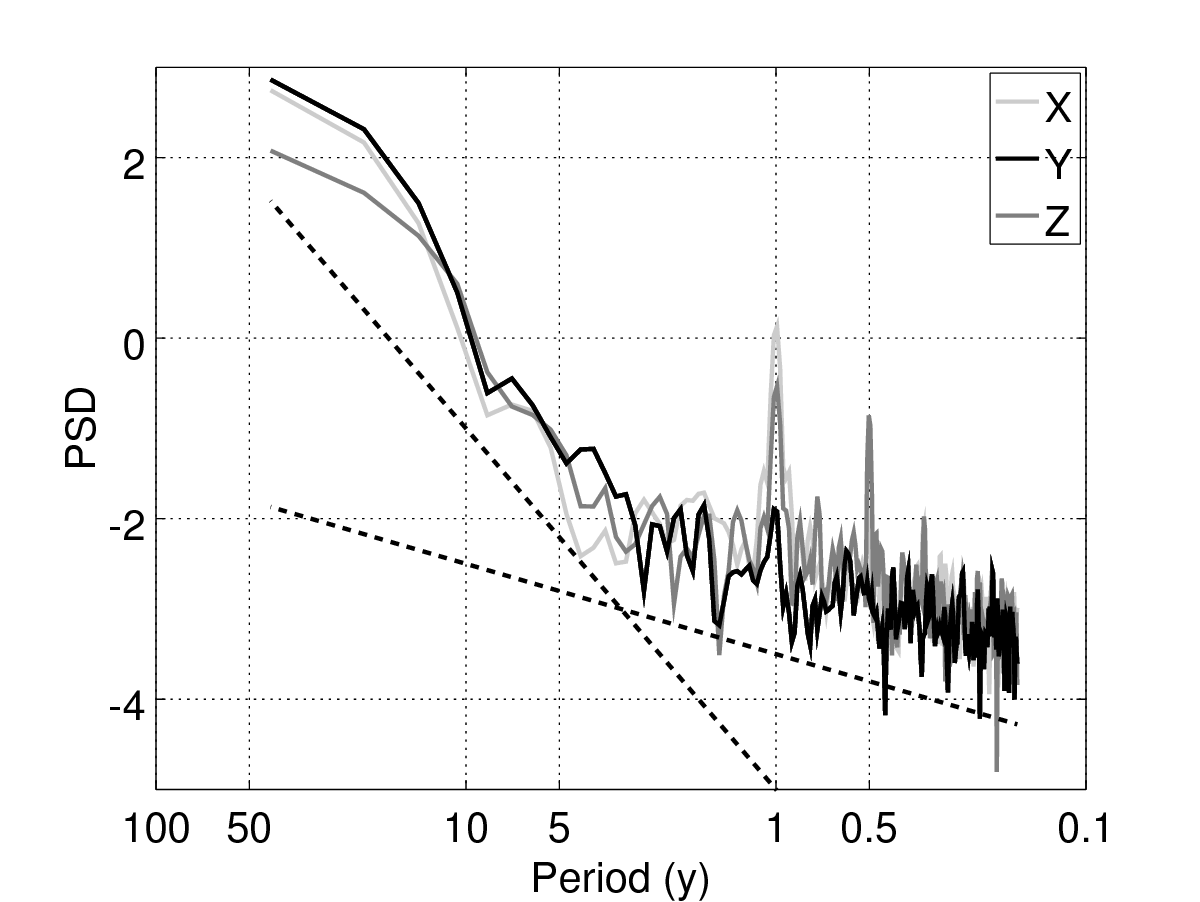}
	\caption{PSD of revised monthly means for the $X,Y$ and $Z$ components at Chambon-la-For\^et ($48^{\circ}$N, $2^{\circ}$E), obtained by a multi-tapper analysis after removing the end-to-end line of the series.
Dotted lines recall the -4 and -1 slopes.}
	\label{fig: PSD XYZ}
\end{figure}

More and more of these features show up as more accurate records accumulate -- uncertainties in observatory series have been significantly reduced with the advent of proton magnetometers in the 1960's, and of digital acquisition later on \citep{turner2015observation}.
Figure \ref{fig: SV changes} shows some examples of SV changes with monthly series at mid, low and high latitudes. 
By removing some estimates of external contributions, the scatter at mid to low latitudes is much reduced, and some of the sudden changes in the SV trend clearly appear only in the cleaned series.
The reduction of the scatter is much less impressive at high latitudes, because there several important external contributions are not satisfactorily modelled.   
The difficult characterization of jerks comes down to the harsh separation of internal and external sources at the very period of interest, from months to a few years. 
This point is highlighted with the power spectral density (PSD) of revised monthly series shown in Figure \ref{fig: PSD XYZ}.  
Even if the dispersion associated with external fields is much reduced at mid-latitudes, remaining external contributions dominate over core signals towards short periods. 
Important spectral lines show up in particular at periods of 1 years, 6 months, etc., which are filtered out when considering annual differences of monthly series as in Figure \ref{fig: SV changes}.

Long standing debates have concerned the community: whether jerks were global or not, delayed or not from one place to the other, etc.
Alternatively, one may wonder whether we should at all localize jerks in time. 
The interpretation of SV series as resulting from processes such that the magnetic field is not twice time differentiable (in a range of frequencies), suggests jerks should naturally emerge at any time \citep{gillet13}. 
This translates from the -4 slope found for the PSD of ground-based series \citep[][and see Figure \ref{fig: PSD XYZ}]{desantis03}, and from a similar slope in the Gauss coefficients PSD \citep{lesur2017frequency}.
In this framework, one should not consider jerks as being characteristic of internal field processes temporally isolated. 
Instead it calls for a physical interpretation of the -4 slope, which is also found in Gauss coefficients series from geodynamo simulations (see \S\ref{sec: time-scales}).

\subsection{On Secular Acceleration Pulses}
\label{sec: SA}
 
The ubiquitous existence of jerks might seem at odds with one focus put forward by the continuous era of satellite measurements: the occurrence of pulses in the norm of the secular acceleration $\partial^2 {\bf B}/\partial t^2$ (SA). 
In satellite field models SA pulses show up every 3 years or so \citep{chulliat2014geomagnetic,finlay2016recent}. 
They seem to occur in between epochs where geomagnetic jerks most likely happen, in which case jerks might not arrive (completely) at any time. 
At the core surface (of radius $c=3485$ km), they are carried by the largest length-scales (degrees $n\le8$), although the SA is less and less constrained towards shorter wave-lengths \citep{lesur2010second}. 
SA pulses were first highlighted in the equatorial belt (within $\pm30^{\circ}$ latitude) in the Atlantic hemisphere, though the more accurate, more recent ones also display important features at high latitudes in the Northern hemisphere. 
A 3 yr period in the SA norm implies 6 yr oscillations in some Gauss coefficients: four SA pusles thus only correspond to two periods, so that longer series are required to assess the exact periodicity of SA pulses, if any, from spectral analyses. 
Similar SA events nevertheless seem to have occurred earlier on, as suggested by \cite{soloviev2017detection} from the analysis of observatory series.

Their interpretation is currently giving rise to many conjectures: MAC waves in stratified layers are candidates \citep{chulliat2015fast}, although SA pulses can entirely be analysed as the signature of quasi-geostrophic (QG) core flows \citep{finlay2016recent}. 
In all cases, fluid motions responsible for such events must contain a significant non-axisymmetric component. 
As a consequence it cannot be the signature of only torsional motions (geostrophic, or  axisymmetric and invariant along the rotation axis, see \S\ref{sec: MC}), which also occur at approximately a 6 yr period \citep{gillet2010fast}. 
It is curious that magnetic data requires intruiguing non-zonal motions at the same period where geostrophic motions are seen to oscillate. 
The possibility of a mechanism coupling these two dynamics is an open question. 

There is nevertheless another possibility for this 3 yr apparent periodicity in SA pulses, in link with (i) the blue SA spatial power spectrum at the CMB and (ii) the way magnetic model are constructed. 
Lets define the SA norm per harmonic degree as a function of time $t$,
\begin{eqnarray}
{\Phi}(n,t)=\sqrt{(n+1)\sum_{m=0}^n \left( {\partial^2_t g_n^m(t)}^2 + {\partial^2_t h_n^m(t)}^2 \right)}\,,
\label{eq: sa time}
\end{eqnarray}
and the SA norm per harmonic degree as a function of frequency $f$,
\begin{eqnarray}
{\Phi^{\dag}}(n,f)=\sqrt{(n+1)\sum_{m=0}^n \left(  \left|{\alpha_n^m}\right|^2(f) + \left|\beta_n^m\right|^2(f) \right)}\,,
\label{eq: sa freq}
\end{eqnarray}
where $\alpha_n^m(f)$ and $\beta_n^m(f)$ stand for the Fourier transform of respectively $\partial^2_t{g}_n^m(t)$ and $\partial^2_t{h}_n^m(t)$. 
From equation (\ref{eq: sa time}) one defines the total SA norm,
\begin{eqnarray}
{\Psi}_N(t)=\sqrt{\sum_{n=1}^N{\Phi}^2(n,t)}\,,
\label{eq: sa norms}
\end{eqnarray}
for a trunctation degree $N$. 
The blue SA power spectrum means ${\Psi}_N$ does not converge as $N$ increases.  

\begin{figure}[t]
\centering
	\includegraphics[width=\linewidth]{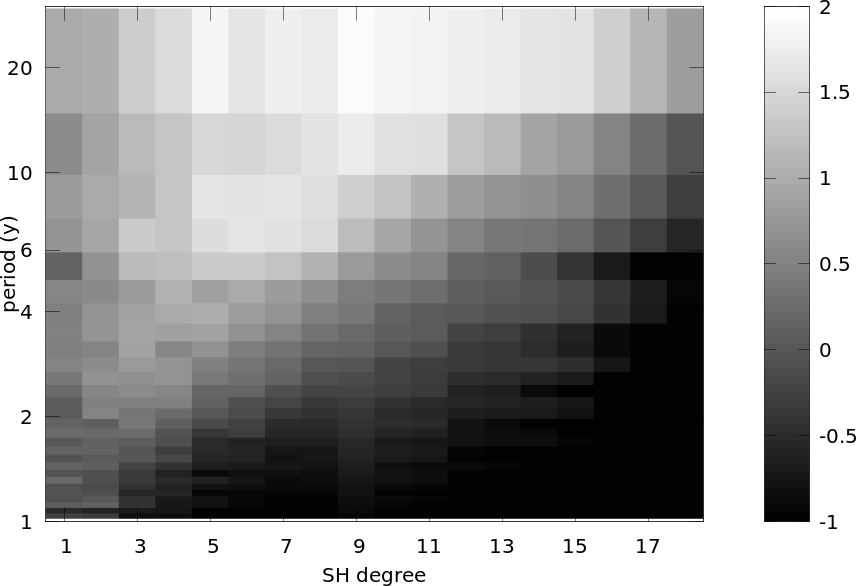}
	\caption{SA norm $\Phi^{\dag}$ as a function of spherical harmonic degree $n$ and period $T=1/f$, as defined in equation (\ref{eq: sa freq}), for the model CHAOS-6-x4  ($\log_{10}$ scale, in nT/yr$^2$). 
}
	\label{fig: SA  norm chart}
\end{figure}

\begin{figure}[t]
\centering
	\includegraphics[width=1\linewidth]{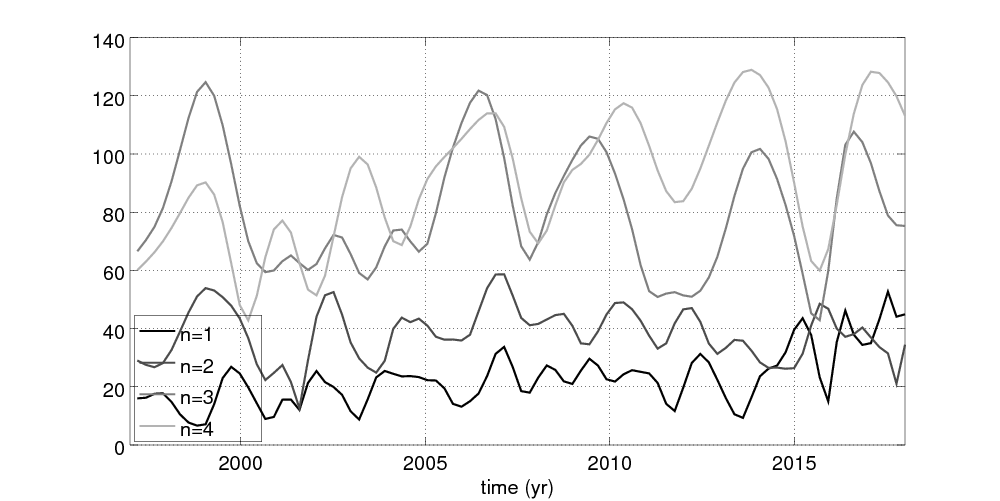}
	\includegraphics[width=1\linewidth]{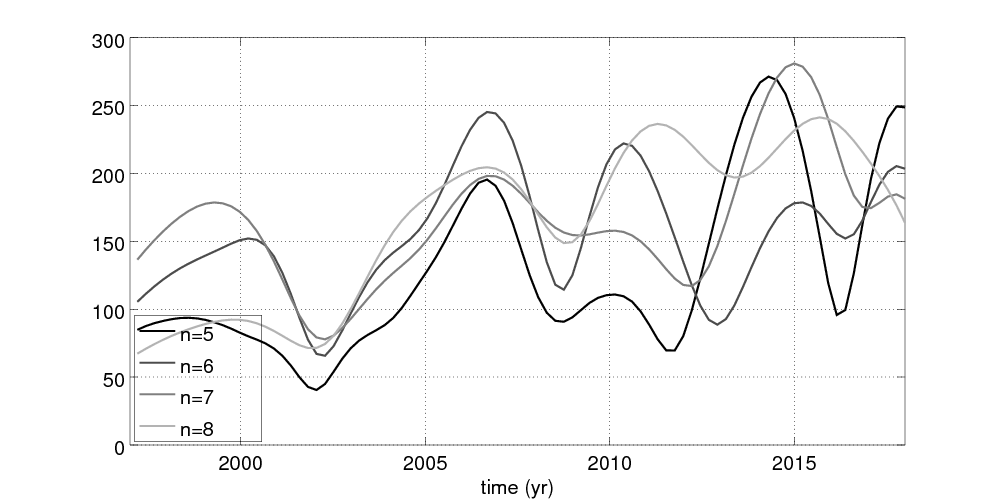}
	\includegraphics[width=1\linewidth]{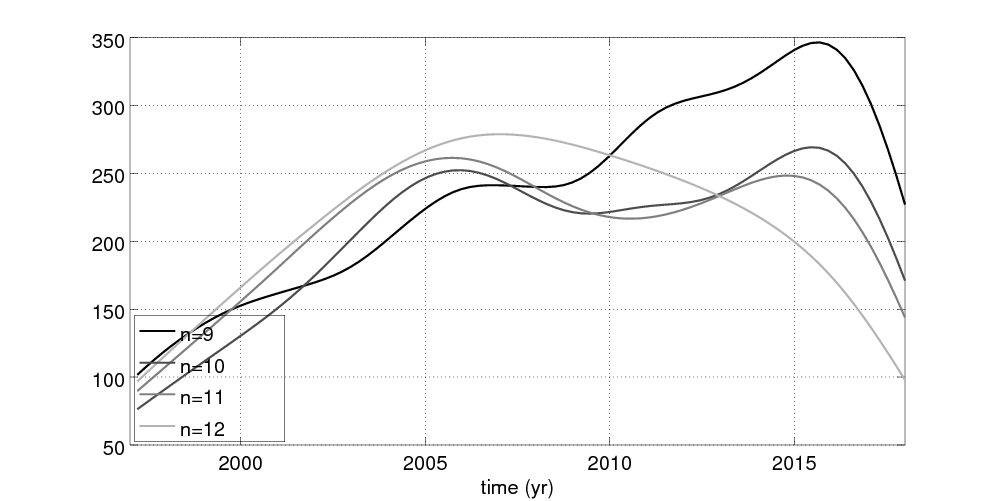}
	\caption{SA norm $\Phi$ as a function of spherical harmonic degree $n$ and time (in nT/yr$^2$, see equation \ref{eq: sa time}), for the model CHAOS-6-x4: degrees $n\in[1,4]$ (top), $n\in[5,8]$ (middle), and $n\in[9,12]$ (bottom).}
	\label{fig: SA  norm per degree}
\end{figure}

Figure  \ref{fig: SA  norm chart} shows ${\Phi^{\dag}}(n,f)$ for the CHAOS-6 field model \citep{finlay2016recent}.
Lower degrees clearly present more power on short time-scales than high degrees, which are more affected by the temporal damping imposed in the model construction. 
At the same time, since at the CMB smaller structures are more energetic, time changes in the total SA norm are dominated by the resolved SV changes at the smaller length-scales, in practice at degrees $n$ from 5 to 8 (see figure \ref{fig: SA  norm per degree}).
This translates into 3 yr SA pulses, coherent with a cut-off period at about 6 yr for the resolution of degrees 5 to 8 in Figure \ref{fig: SA  norm chart}. 
For $n\ge 9$, interannual SV changes are not recovered, and these smaller length-scales do not imprint SA pulses, while at large length-scales (in particular $n\le 2$) the SA norm oscillates at periods shorter than 3 yr.  

\begin{figure}[t]
\centering
	\includegraphics[width=\linewidth]{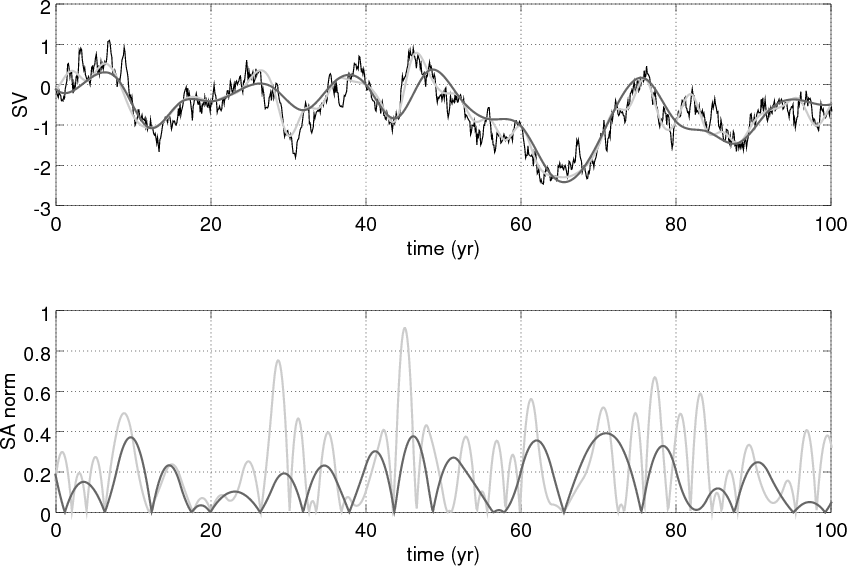}
	\caption{Top: 1D synthetic `SV' time series $d\varphi/dt$ (black) generated as an AR-1 process, with a correlation time of 10 yr and a unit variance: the series are interpollated using splines with a sampling of respectively 2 yr (light grey) and 4 yr (dark grey). 
Bottom: `SA' norm series $|d^2\varphi/dt^2|$ evaluated for the two interpollated series. 
}
	\label{fig: SA  norm synthetic}
\end{figure}

As an illustration, let consider a synthetic one-dimensional `SV' series $d\varphi/dt$ generated as an order 1 auto-regressif process (AR-1). 
Such a process naturally shows jerk-like feature (figure  \ref{fig: SA  norm synthetic}, top). 
By construction, it is not differentiable: the spectrum for its time derivative $d^2\varphi/dt^2$ (or `SA')  would be white, i.e. pulses in the SA norm $\displaystyle |d^2\varphi/dt^2|$ occur all the time, with an amplitude depending on the sampling rate.
Considering now the same series but low-pass filtered, we see oscillations in the `SA norm' of period half the shortest non-filtered time-scale (figure \ref{fig: SA  norm synthetic}, bottom).
This test does not mean there is no signal associated with reported SA pulses. 
These sign interannual SV changes at intermediate length-scales, whose phase appears coherent within several degrees at a period about 6 yr (Figure \ref{fig: SA  norm per degree}), in link with the localized patches isolated in field models.
However, one should keep in mind that SA pulses should not necessarily be interpreted as resulting from periodic processes, acknowledging our inability to recover rapid fluctuations at small length-scales. 

Finally, SV and SA highlight the core dynamics at different frequencies (time derivatives enhance short periods). 
Given the red temporal spectrum observed in magnetic series at decadal to centennial periods  \citep[of slope about respectively -4 and -2, see][]{desantis03,panovska2013observed}, the SV is dominated by decadal or longer fluctuations 
while the SA emphasizes the shorter accessible time-changes (a couple of years). 
The question of the continuation of the -4 temporal spectrum towards shorter periods, and a possible cut-off frequency in link with the mantle conductivity or the magnetic dissipation in the fluid core, remains open. 

\section[]{Dynamical Models Of Earth's Core Dynamics}
\label{sec: dynamics}

\subsection{Relevant Time-Scales}
\label{sec: time-scales}

Minimalists models of the dynamics within the Earth's core must include, on top of the induction equation (\ref{eq: induction}), the momentum equation in the rotating frame of our planet,
\begin{multline}
\frac{\mathrm{D}{\bf u}}{\mathrm{D}t}+2\Omega{\bf 1}_z\times{\bf u}=-\nabla P\\
+ \frac{1}{\rho\mu}\left(\nabla\times{\bf B}\right)\times{\bf B}+{\bf F}+\nu\nabla^2{\bf u}\,.
\label{eq: momentum}
\end{multline}
${\bf 1}_z$ is the unit vector along the rotation axis, $\Omega=7.27\,10^{-5}$ rad/s is the rotation rate, $P$ is the modified pressure (including the centrifugal force), $\nu\sim10^{-5}$ m$^2$/s is the kinematic viscosity, $\mu=4\pi\,10^{-7}$ H/m the magnetic permeability of free space, $\rho\simeq10^4$ kg/m$^3$ the core density, and ${\bf F}$ contains body forces (such as buoyancy). 
Several characteristic time-scales can be extracted from equations (\ref{eq: induction}) and (\ref{eq: momentum}), 
\begin{itemize}
\item the vortex turn-over time $\tau_u=L/U$, for a circulation of size $L$ at a flow speed $U$,
\item the magnetic diffusion time $\tau_{\eta}=L^2/\eta$,
\item the time $\tau_{a}=D/V_a$ taken for a perturbation to cross the core, at the Alfv\'en speed $V_a=B/\sqrt{\rho\mu}$, with $B$ the magnetic field magnitude and $D=2265$ km the gap between the inner and outer cores,
\item the time $\tau_i=D/(\Omega L)$ (characteristic of inertial waves) taken for a perturbation of size $L$ to form a Taylor column of height $H\sim D$ \citep{cardin2015experiments}.
\end{itemize}
I refer to \cite{nataf2015turbulence} for a discussion of time-scales and spatio-temporal spectra in planetary cores. 
For the sake of simplicity, let consider planetary size structures ($L\sim D$), $U(D)\simeq10$ km/yr, characteristic of the westward drift at the core surface \citep{finlay03}, and $B\sim 3$ mT for the field strength within the core \citep{gillet2010fast}. 
From these we obtain
\begin{multline}
\tau_{\eta}\sim 100,000 \,\mathrm{ yr} \gg \tau_u\sim 200\, \mathrm{ yr}\\ \gg \tau_a\sim 2.5\, \mathrm{ yr} \gg \tau_i\sim 4\,10^{-4} \,\mathrm{ yr}\,.
\label{eq: times}
\end{multline}

I consider below convective dynamos that couple equations (\ref{eq: induction}--\ref{eq: momentum}), where ${\bf F}$ stands for buoyancy, with a conservation equation for heat (often in a codensity formulation that mixes thermal and chemical density anomalies with a single diffusivity). 
From the above time-scales one can build several dimensionless numbers of interest, summarized in Table \ref{tab: numbers} for the Earth and numerical geodynamos. 
I focus on dynamos targetting Earth-like parameters (low Ekman and magnetic Prandtl numbers). 
For these two input parameters, simulations are still far away from geophysical values. 
Interestingly, ratio of magnetic diffusion and induction (the magnetic Reynolds number) are nevertheless correctly recovered by numerical studies. 
The Earth's core is luckylly close to dynamo onset, as three-dimensional computations would struggle to reach significantly larger values of $R_m$. 
The ratio of inertal to Alfv\'en time-scales also is well recovered (as measured by a Lehnert number $\lambda\ll 1$).  
This transpires through a rotation dominated dynamics on short periods (quasi-geostrophy): motions organize primarily in columns aligned with the rotation axis \citep{jault2008axial,soderlund2012influence}.

To be compared with geophysical observations, simulations must be dimensionalized, a procedure that involves a priori  choices. 
One often resort to the magnetic diffusion time when long sequences ($>O(10)$ kyr) are to be considered, while focusing on rapid changes a scaling based on the advective turn-over time appears efficient \citep{olson12}. 
Interestingly, $\tau_u$ in simulations is connected to the time-scale $\tau_n$ ratio of the MF to SV spatial power spectra.
Indeed, \cite{lhuillier2011earth} approximate $\tau_n\simeq 3\tau_{u}/n$.
Measurable from geomagnetic field models, $\tau_n$ can thus be used to compare geophysical and numerical outputs. 

Focusing on periods from decades up to some 10 kyr, dynamo simulations share some spectral properties with observations. 
A -4 slope of Gauss coefficients temporal spectra is found at decadal to centennial periods (see \S\ref{sec: jerks}), and a transition from a -4 to a -2 slope towards long periods is observed for the axial dipole \citep{olson12,bouligand2016frequency,buffett2015power} as well as in geophysical records of the virtual axial dipole moment \citep{constable05}. 
However, what is missed by computations is actually low values of the Alfv\'en number, which means all time-scales between $\tau_a$ and $\tau_u$ (a few years to a few hundreds of years) are at least partly shrunk: while the magnetic energy is $10^4$ times larger than the kinetic energy at large length-scales in the Earth's core, both share similar magnitudes in simulations.  
The good correspondance in PSD highlighted above should not hide the too narrow separation of time-scales, consequence of enhanced diffusive processes \citep{bouligand2016frequency}.

\begin{table*}
 \caption{Several dimensionless numbers, estimated for the Earth's core and for numerical simulations of the geodynamo \citep[adapted from][]{schaeffer2017turbulent}.}
\centering
 \begin{tabular*}{1\hsize}{@{\extracolsep{\fill}}|l|rrrrrr|}
\hline\noalign{\smallskip}
  & $P_m$ & $E$ & $R_m$ &  $A$ & $S$ & $\lambda$ \\
\hline\noalign{\smallskip}
	 &  magnetic &  &  magnetic &  &  & \\
Name &  Prandtl & Ekman &  Reynolds & Alfv\'en & Lundquist & Lehnert\\
\hline\noalign{\smallskip}
Definition & $\displaystyle \frac{\nu}{\eta}$ & $\displaystyle \frac{\nu}{\Omega D^2}$ & 
$\displaystyle  \frac{\tau_{\eta}}{\tau_u}=\frac{UD}{\eta}$ & $\displaystyle \frac{\tau_{a}}{\tau_u}=\frac{U}{V_a}$ & $\displaystyle \frac{\tau_{\eta}}{\tau_a}=\frac{V_aD}{\eta}$ & $\displaystyle \frac{\tau_{i}}{\tau_a}=\frac{V_a}{\Omega D}$\\
\hline\noalign{\smallskip}
Earth & $10^{-5}$ & $10^{-14}$ & $10^3$  & $10^{-2}$ & $4\,10^4$ & $10^{-4}$\\
simulations & $>10^{-1}$ & $>10^{-7}$ & $10^3$ & $>3\,10^{-1}$ & $<10^3$ & $>10^{-4}$ \\
\hline\noalign{\smallskip}
 \end{tabular*}
 \label{tab: numbers}
\end{table*}

Reaching weaker values of $A$ is actually computationnally very expensive, because it implies increasing both the rotation rate and the input thermal forcing \citep{schaeffer2017turbulent}. 
The alternative is two-folded. 
One avenue consists of parameterizing turbulent magneto-hydro-dynamic (MHD) processes, as performed by \cite{aubert2017spherical}. 
Enhancing dissipation for large wave numbers, the authors manage to follow a path in direction of Earth-like parameters. 
Alternatively, one may consider dynamos for $P_m\ge1$ \citep{kageyama2008formation}, so that the strong magnetic energy branch is captured \citep{dormy2017three}. 
Such dynamos nevertheless show much less axial invariance, as they reach significantly larger values of $\lambda$. 
Furthermore, what they bring is at the expense of losing wave-dominated rapid dynamics, by enhancing the role played by viscous processes.

These limits call today for considering alternative reduced equation systems. 
I discuss below two families of dynamical models relevant for interpreting magnetic data: 
magnetostrophic waves (\S\ref{sec: MC}) where Coriolis and Lorentz forces play the dominant roles, and MAC waves (\S\ref{sec: MAC}) that occur in the presence of a stratified layer at the top of the core.

\subsection{Magnetostrophic Waves and Taylor's State}
\label{sec: MC}

Major developements have been carried out ignoring at first sight the source term in the momentum equation. 
Viscosity too shall be neglected in this section, on the argument $E\ll 1$.
Remain two coupled equations (induction and momentum) giving birth to slow and fast MC modes \citep{malkus1967hydromagnetic}, with restoring mechanisms from Coriolis and Lorentz forces.
Noting $\omega_i=O(2\Omega)$ the frequency of inertial waves (anisotrope and dispersive) and $\omega_a \sim V_a/L$ that of Alfv\'en waves (non dispersive), the frequencies of fast and slow MC waves evolve respectively as \citep{finlay2008course}
\begin{eqnarray}
\omega_{MC}^{f}\sim \omega_i\;\;,\;\;
\omega_{MC}^s\sim \omega_a^2/\omega_i\,.
\label{eq: MC}
\end{eqnarray}
If rapid MC modes are essentially inertial waves, too fast to be detected today from geomagnetic data, the period of the gravest slow ones is $\tau_{MC}^s\sim 2\Omega L^2/V_a=O(500)$ yrs for $L\sim 2000$ km. 
They are candidate to interprete major features such as westward drift of equatorial patches, or fluctuations found in archeomagnetic series (see \S\ref{sec: long periods}). 

Slow MC waves are magnetostrophic, in the sense that inertia is negligeable. 
They thus satisfy to \cite{taylor1963magneto}'s condition, 
\begin{eqnarray}
\forall s, {\bf 1}_{\phi}\cdot\iint_{\Sigma}\nabla\times{\bf B}\times {\bf B}\mathrm{d}\Sigma=0 \,,
\label{eq: taylor}
\end{eqnarray}
obtained by integrating equation (\ref{eq: momentum}) along geostrophic cylinders $\Sigma(s)$ (encapsulated in the outer core, and which axis coincides with the rotation axis), with $s$ the cylindrical radius and ${\bf 1}_{\phi}$ the unit vector in the azimuthal direction. 
Said differently, in absence of inertia and viscosity, since the projections of the Coriolis force and buoyancy on geostrophic cylinders vanish, the only remaining term comes from the Lorentz force. 
Numerical attempts at reaching magnetostrophic dynamos have been carried out \citep{livermore2011evolution,roberts2014modified}, involving challenging computational issues \citep{walker1998note}.

Magnetostrophic waves contain a sub-family, QG-MC modes \citep{hide1966free}. 
The strong axial invariance of flows found in rotating MHD simulations gave birth to a renewed interest in such waves \citep{canet2014hydromagnetic}. 
Such modal analyses constitute a first step towards reduced models advecting quadratic quantities of the magnetic field in the equatorial plane \citep{canet2009forward,jault2015waves}, where the main sources of nonlinearities arise from the electro-motive force in equation (\ref{eq: induction}). 
The time-stepping of such equations is in its early phase \citep{maffei2017kinematic}.

Interestingly, nonlinear interactions of (QG-)MC modes are able to fill the entire frequency spectrum from interannual to centennial periods, where the most accurate magnetic observations are available.
However, similar intermediate time-scales also emerge from the description of hybrid inertial--Alfv\'en waves by \cite{bardsley2016inertial}: perturbations propagate along ${\bf 1}_{z}$ to form columns that travel along field lines at the (local) Alfv\'en speed.

The validity of the QG assumption is nevertheless debated. 
Kinematic core flow inversions, applied to the most recent era covered by satellites, require a core surface kinetic energy dominated by equatorially symmetric structures \citep{gillet11,aubert2014earth}.
This observation is nevertheless tempered by the aspiration for the flow to locally cross the equator, in particular under Indonesia  \citep{bloxham1989simple,baerenzung2016flow,barrois2017contributions} where some of the most intense SV activity is detected. 
Dynamo simulations too require some breaking of axial invariance for Earth-like dipolarity to be obtained \citep{garcia2017equatorial}, and deviations from non-magnetic QG solutions are also found in linear magneto-convection solutions with an imposed dipole field \citep{sreenivasan2011helicity}.
This motivates the derivation of baroclinic QG models, where motions, still strongly anisotrope, are not axially invariant \citep{calkins2013three}.

\begin{figure}
\centering
	\includegraphics[width=1\linewidth]{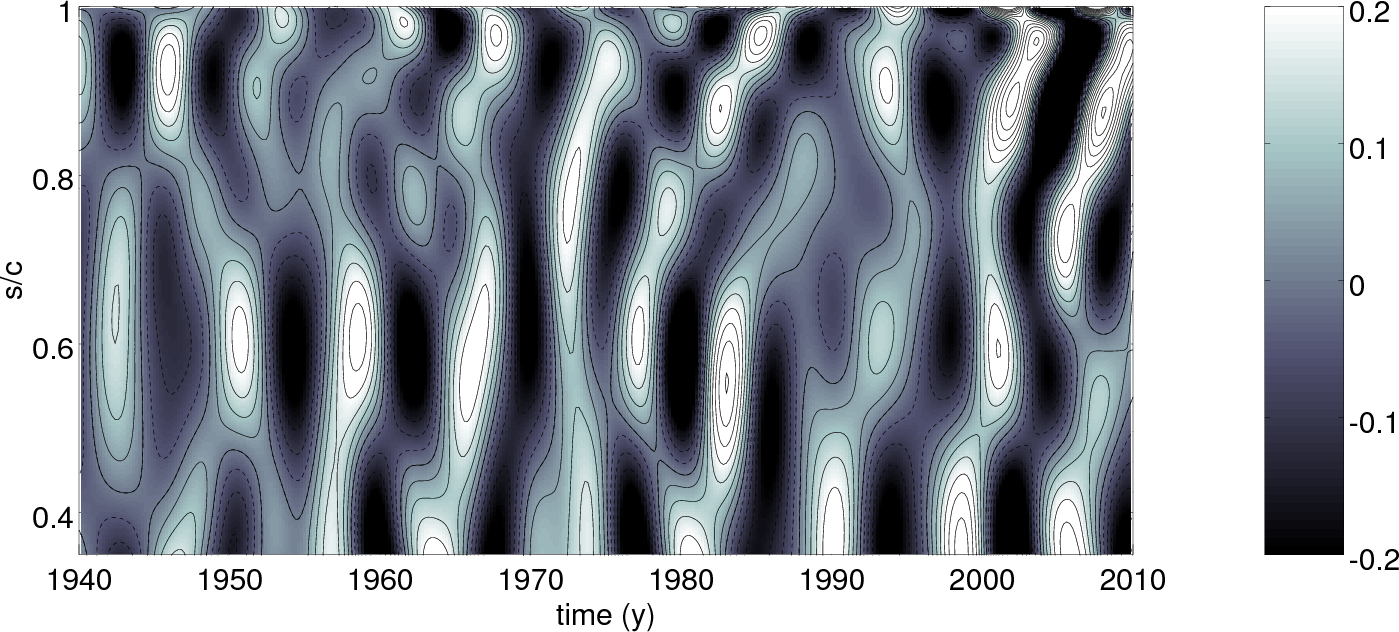}\\
	\includegraphics[width=1\linewidth]{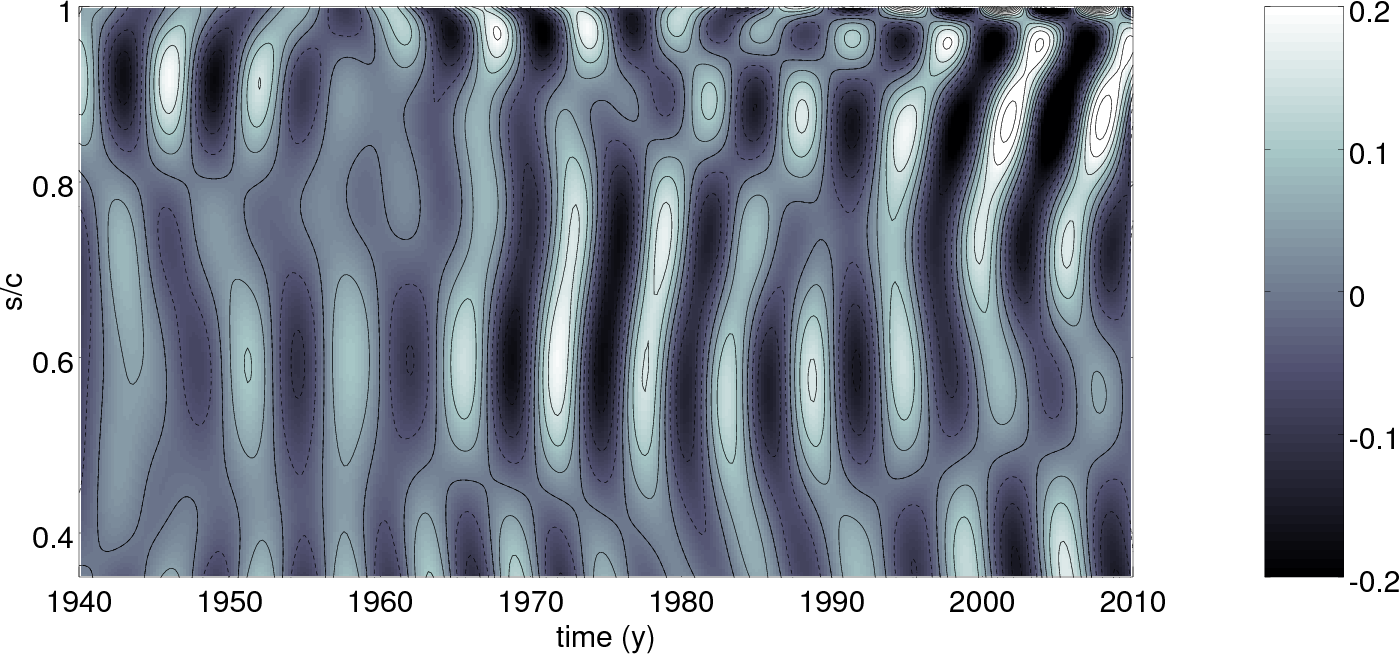}
	\caption{Geostrophic velocity as a function of time and cylindrical radius (in km/yr), between the inner core radius ($s/c=0.35$) and the core equator ($s/c=1$). 
From kinematic QG inversion \citep{gillet2015planetary} band-pass filtered for $T\in[4, 9.5]$ yrs (top) and $T\in[5, 8]$ yrs (bottom).}
	\label{fig: TW}
\end{figure}

A specific family of QG motions has been the focus of many studies since their theoretical description by \cite{braginski70}: torsional (or azimuthal) motions of geostrophic cylinders.
These arise when re-instating inertia in equation (\ref{eq: taylor}): any perturbation to Taylor's constraint shall give rise to a wave-like response. 
Because inertial waves are much faster than Alfv\'en waves at large length-scales, torsional waves are actually the only Alfv\'en waves detectable in the Earth's core (as the Coriolis force averaged over geostrophic cylinders vanishes). 
These waves have been detected from magnetic observations, propagating outward from the inner core \citep{gillet2010fast}. 
Their detection offered the possibility to probe the magnetic field in the bulk of the core. 
Their excitation mechanism is a source of debate:  
\cite{teed2015transition} argue from magneto-convection simulations for instabilities related to large Lorentz torques on the edge of the tangent cylinder (the geostrophic cylinder tangent to the inner core), while \cite{gillet2017excitation} consider the outward propagation as the signature of propagating normal modes in the presence of a conductive layer at the base of the mantle, triggered either on the tangent cylinder or in the bulk of the fluid core. 
Band-pass filtered geostrophic motions inferred from magnetic data (Figure \ref{fig: TW}) show waves modulated over decadal periods, though it is difficult to accurately isolate spectral lines (and thus the wave forms), given (i) the limited time-span covered by accurate observations, (ii) the blurred SV signal towards high frequencies (see \S\ref{sec: SA}), and (iii) the red spectrum of core motions, which are less energetic towards short periods. 

Their apparent partial absorption at the equator is a potential source of information on the deep mantle conductivity (hardly detectable from above) in a case where an electro-magnetic torque couples the core to the mantle \citep{schaeffer2016electrical}. 
In this context, mantle conductances deduced from torsional waves \citep[$3\,10^7-10^8$ S,][]{gillet2017excitation} appear relatively low in comparison with previous estimates from either core nutations \citep{buffett2002modeling} or kinematic core flows \citep{holme1998electromagnetic}.
Spread over a thick ($\sim1000$ km) shell, such values remain nevertheless higher than what is currently accepted from induction studies \citep[typically between 0 and 10 S/m,][]{velimsky2010electrical}. 
Remain several possibilities: 
(i) materials of electrical conductivity intermediate between $O(10^2)$ S/m and that of the core ($10^5$ S/m) may be shrunk into a thin layer close to the CMB (to which induction studies are insensitive), 
(ii) the dissipation in models of torsional waves may be under-estimated, or 
(iii) the core-mantle coupling could arise from another mechanism than electro-magnetic torques (e.g. topographic).

Another motivation for the study of geostrophic motions is their link to changes in the length-of-day \citep[LOD,][]{jault2015waves}. 
These show modulated fluctuations of typically 0.2 ms around 6 yr periods \citep{chao2014earth} convincingly predicted by geostrophic flow models inverted from magnetic data \citep{gillet2015planetary}.
In such a scenario, the larger LOD changes at decadal periods (of amplitude a few ms) would result from the interaction of non-geostrophic motions with the background magnetic field (as derived from Taylor's condition), though an alternative scenario is reported below. 
At periods decadal and longer, the dynamics of both non-geostrophic and geostrophic motions could be associated with MC waves, as these potentially couple zonal to non-zonal motions in the presence of non-axisymmetric background magnetic fields \citep{labbe2015magnetostrophic}.

\subsection{MAC Waves In A Stratified Layer}
\label{sec: MAC}

Motivated by the detection of a low seismic velocity layer at the top of the core \citep[some 100 km thick,][]{helffrich2010outer}, an extra complexity may be introduced. 
As the inner core cristallizes, the fluid phase is enriched in light elements that potentially accumulate at the top of the core  \citep[a point that nevertheless requires elaborate scenarii, see][]{brodholt2017composition}.
Low seismic velocities can be translated into density gradient. 
The re-evaluation to larger values of the outer core thermal conductivity from high temperature, high pressure ab initio calculations and laboratory experiments \citep[e.g.][]{pozzo12,ohta2016experimental}, also call for a subadiabatic heat flux and a stable layer at the top of the core \citep[but see][]{konopkova2016}. 

A resulting stable stratification would carry gravity waves, of frequency $\omega_g$ characterized by the Br\"unt-Va\"isala frequency $\displaystyle N=\sqrt{-\frac{g}{\rho}\frac{d\rho}{dr}}$, where $g$ is the gravity acceleration.
The penetration of geostrophic columns through such a layer depends on the relative periods of inertial and gravity waves \citep{takehiro2001penetration,vidal2015quasi}, which much vary with the considered length-scales. 
In the Earth's core we do not know accurately how $N$ compares with $\Omega$. 
In the case of QG inertial (i.e. Rossby) waves, the shorter the length-scale, the longer the period, so that large length-scale modes (the one we may see) are less likely to be affected by stratification. 
Interestingly, most nodes of such waves concentrate near the equator, where intense SA pulses have been first highlighted (see \S\ref{sec: SA}). 

\cite{vidal2015quasi} found weak effect of a simple geometry imposed magnetic field on the penetration depth for Rossby modes. 
On the other end of the time spectrum (towards zero frequency), \cite{takehiro2015penetration} and \cite{takehiro2017penetration} show that the presence a magnetic field couples the stratified layer to the deeper buoyant layer, if the radial propagation of Alfv\'en waves along poloidal field lines is faster than their damping by diffusive processes. 
Applied to the Earth, the strong value of the Lundquist number $S$ implies deep convective motions to penetrate into the stable layer. 
If dynamo simulations with a top stratified layer show a lower level of SV at the CMB \citep{christensen2008models}, it is because of relatively low values of $S$, which may apply to planets like Mercury (possibly Saturn).
An alternative scenario for reconciling magnetic and seismic observations involves lateral heat flux heterogeneities: locally buoyant areas may lay within a globally stable density gradient below the CMB \citep{olson2017dynamo}. 
We still lack definitive arguments on the influence of stratification at decadal periods: the penetration depth of slow QG-MC waves into a stratified layer, or their coupling with MAC modes in the top layer, require dedicated studies. 

Meanwhile, in absence of thermo-chemical diffusion, and under the Boussinesq approximation, the system may be reduced following \cite{si1993mac}'s hidden ocean model. 
In the presence of a magnetic field, the linearized system of equations leads to MAC waves within the stably stratified layer. 
Their period evolves as \citep{finlay2008course}
\begin{eqnarray}
\omega_{MAC}=\pm\omega_{MC}^s\left(1+\frac{\omega_g^2}{\omega_a^2}\right)^{1/2}\,,
\label{eq: MAC}
\end{eqnarray}
which applied to Earth-like values ($\omega_a\ll\omega_g$) gives $\omega_{MAC}\simeq\pm\omega_a\omega_g/\omega_i$. 
Associated periods span interannual to decadal time-scales, the reason why they have been considered to explain features such as SA pulses \citep{chulliat2015fast} or fluctuations in the axial dipole SV \citep{buffett2014geomagnetic}. 

Interestingly, if zonal motions are not directly affected by stratification (they do not cut iso-density surfaces), their coupling to non-zonal motion may render them sensitive to the density gradient.
As a consequence there exist axisymmetric MAC waves  \citep{braginsky1999dynamics}, which carry angular momentum and thus a signature in the LOD. 
Recent numerical estimates however suggest a too small angular momentum budget in the upper layer \citep{buffett2016evidence}, which would call for a magnetic coupling with geostrophic motions in the deeper layer. 

\subsection{Insights From Numerical Geodynamo Simulations}
\label{sec: dynamos}

There is no such dynamo computation suited to cover the entire frequency range \citep{meduri2016simple}. 
Dynamos whose parameters $E$ and $P_m$ are the closer to Earth-like values do not show polarity reversals -- though it is not possible to intergrate primitive equations for long enough to likely see a reversal with extreme parameters. 
Reversals are reserved to simulations with high $E$, for which forcings strong enough to produce reversals are easier to compute.
On the contrary, torsional waves excited by Lorentz torques require fast enough rotation rates \citep{teed2015transition}, and they are continuously excited for the lowermost values of $P_m$ and $E$ \citep{schaeffer2017turbulent}.  
As another example, MC waves can be detected in simulations, provided a strong enough field is generated \citep[on the strong branch of][]{dormy2017three}, which calls for using either large $P_m$ or low $E$ \citep{hori2017dynamics}. 

\begin{figure}
\centerline{
	\includegraphics[width=.8\linewidth]{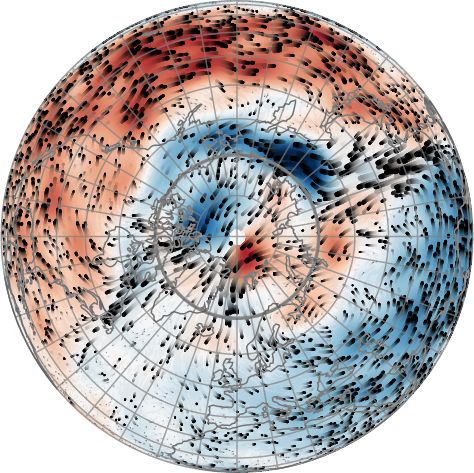}}
\centerline{
	\includegraphics[width=.8\linewidth]{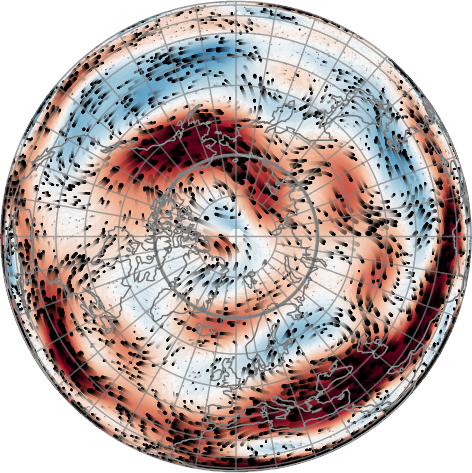}}
\centerline{
	\includegraphics[width=.8\linewidth]{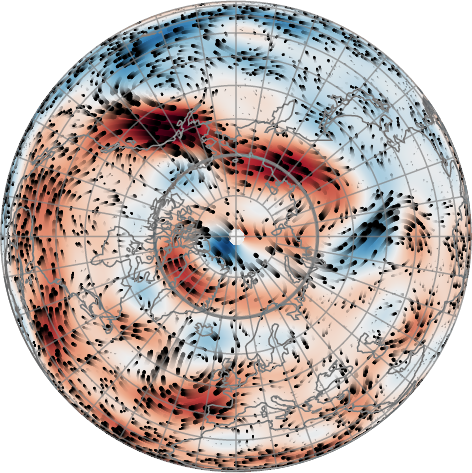}}
	\caption{
Equatorial cross-sections of core motions, viewed from the North pole: passive tracer trajectories (black) superimposed with $u_{\phi}$ (colorscale within $\pm 30$ km/yr, red westward).
Top: geodynamo simulation with isotropic forcing \citep[$E=10^{-6}$, $P_m=0.2$,][]{schaeffer2017turbulent}, low-pass filtered for $T\ge 1000$ yrs.  
Middle: kinematic QG inversion in 2008 \citep{gillet2015planetary}. 
Bottom: geodynamo simulation with non-isotropic forcing favoring a westward drift in the Atlantic hemisphere \citep[$E=1.2\,10^{-5}$, $P_m=2.5$,][]{aubert2013bottom}.
}
	\label{fig: gyre}
\end{figure}

Dynamo simulations show today the first signs for the physics thought to occur in the Earth's core and probed by direct magnetic records. 
It is also an extensive tool to tackle issues such as the core-mantle coupling. 
As an example the westward drift and the asymmetry between Atlantic and Pacific hemispheres found in modern SV maps is interpreted by \cite{aubert2013bottom} as the by-product of a gravitationnal torque between the inner core and the mantle, in link with tomographic seismic anomaly maps.
Alternatively to the thermally driven asymmetries and persistent flux lobes \citep[e.g.][]{gubbins2007correlation,aubert2008thermochemical}, the rich dynamics operating in the most up-to-date dynamo simulations with isotropic forcing starts to show some breaking of the longitudinal symmetry naturally emerging. 
\cite{schaeffer2017turbulent} indeed extracted almost axially invariant eccentric circulations persistent over centuries, similar to that inverted by \cite{pais2008quasi} from satellite field models. 
A comparison of eccentric circulations found from magnetic data and the two above computations is shown in figure \ref{fig: gyre}. 
The origin of large length-scale circulations in computations is subject to discussion: \cite{sakuraba2009generation} advocates for an enlarging effect in link with the fixed-flux thermal boundary condition, while \cite{yadav2016approaching} relates the energy cascade towards wider structures to highly supercritical convection allowing to reach a MAC equilibrium of forces (and a stronger effect of the magnetic field).
\cite{schaeffer2017turbulent} also finds wider patterns where the field is stronger. 

\section[]{The Advent Of Geomagnetic Data Assimilation}
\label{sec: assimilation}

The past decade has seen rising geomagnetic data assimilation studies. 
These aim at mixing information from observations and from a dynamical model to constrain the state of the Earth's core -- a discrete representation of unknown fields, for instance ${\bf u}$, ${\bf B}$...  \citep[for a review in the context of geomagnetism, see][]{fournier2010introduction}.
As such, they depart from kinematic core flow inversions. 
The entry to the fluid core is made via the radial component of the magnetic field at the core surface. 
This latter, continuous through the CMB when the mantle is insulating, is obtained by downward continuation of a potential field from magnetic records above the Earth's surface (see \S\ref{sec: separation}). 

The main assimilation techniques essentially split between variational and sequential approaches.
With the former one adjusts the initial condition
to minimize the misfit to the data over the whole considered time window, subject to the numerical constraint imposed by means of adjoint equations. 
The latter sequential avenue has been more widely used, especially when the forward model consists of geodynamo equations, as initiated by \cite{liu2007observing}. 
It consists in operating a series of analyses (a linear regression given the propagated model error covariance matrix, ${\sf P}^f(t)$) and forecasts (advection of the core state by the model) each time observations are available.
I recall below the several attempts that have been carried out so far, with either self-consistent geodynamo simulations, or reduced models. 
The pertinence of the considered model will depend on the time-scale of interest. 

\subsection{Snapshot Inference Using Geodynamo Norms}
\label{sec: snapshot}

The most up-to-date simulations (with the lower values of $A$ and $E$) are not accessible yet for assimilation studies, as ensemble strategies followed to approximate ${\sf P}^f$ require to compute simultaneously several tenths (or more) of geodynamo runs (prohibitive using cutting-edge computations). 
As a consequence, progresses are still needed for directly adjusting the trajectory of dynamo simulations using modern observations and their interannual to decadal fluctuations (see \S\ref{sec: SA}). 

Such computations nevertheless provide Earth-like snapshot images of the core surface field \citep{christensen2010conditions}.
This encouraged the combination of spatial cross-covariances, obtained from long geodynamo free runs, with MF and SV observations, into snapshot images of the core state \citep{fournier2011inference}.
In this framework, observing SV data (on the top of MF data) helps constrain unobserved quantities such as core motions, via the radial component of equation (\ref{eq: induction}), when the dynamics of the system is either ignored on purpose \citep{aubert2014earth} or biased due to input parameters far from realistic \citep{kuang2015dynamic}. 
This comes down to inferring the core state in a kinematic manner, using a dynamo norm as a spatial prior. 

Core flow maps obtained with such dynamo norms recover the planetary scale eccentric gyre first revealed with kinematic QG inversion \citep{pais2008quasi}. 
They also give partial access to the underlying magnetic field and co-density organisation \citep{aubert2013bottom,aubert2014earth}. 
These latter two fields are indirectly related to surface MF and SV observations, through their statistical cross-correlation with the large length-scales magnetic and velocity fields at the core surface. 

\subsection{Geodynamo Driven Dynamical Reconstructions}
\label{sec: assim dynamo}

Geodynamo simulations are arguably suitable for dynamically inferring changes in the core state on centennial and longer periods. 
This paves the way for assimilation of historical and archeomagnetic data, implementing time constraints from primitive equations. 
The first attempts \citep{kuang2009constraining} have been performed with `optimal interpolation' (OI) sequential techniques (i.e. with a static ${\sf P}^f$), and univariate statistics (ignoring most cross-covariances), in order to reduce the computational cost.   
It is however known that space and time cross-correlations are important \citep{fournier2011inference,tangborn2015geodynamo}.

In this context, \cite{fournier13} used an ensemble Kalman filter (EnKF).
Updating the forecast statistics was possible thanks to the limited amount of data, 
in practice $n_b(n_b+2)$ with $n_b$ the spherical harmonic truncation degree when one considers field models as observations ($n_b\simeq 5$ and $13$ for respectively archeomagnetic and historical models). 
Using synthetic experiments at relatively high values of $E$, \citeauthor{fournier13} show it is possible to retrieve some structures of the unobserved quantities (such as the field and the flow in the bulk of the core) from surface observations of $B_r$.  
They advocate for ensemble sizes significantly larger than the size of the data vector, in order to reduce the warm-up time to about 1000 yrs, and thus significantly supplement OI techniques that do not update ${\sf P}^f$. 

\cite{sanchez2016phd} recently extended this approach to the re-analysis of point-wise (synthetic) archeomagnetic observations at the Earth's surface. 
Similarly to \cite{fournier13}, she finds possible to recover part of the un-observed state even when data are mostly concentrated in a single hemisphere. 
This is encouraging: by directly inverting for ancient data one avoids ad hoc assumptions made to recover Gauss coefficients and their associated uncertainties (see \S\ref{sec: uncertainties}). 
It nevertheless requires the migration in time of data at the precise epochs where analyses are performed. 
Alternatively, one may use uncertainties as provided by stochastic inversions \citep{hellio18}. 
These works pave the way for applications to geophysical data, though it certainly requires a closer look at errors associated with the dynamical model itself.  

This last issue is two-folded. 
First, rapid changes as captured with modern data are not reproduced yet with primitive equations based models: the model is unperfect and should be complemented. 
Second, the observation is very sparse: we have access only to the large length-scales of $B_r$ at the top of the fluid core. 
This induces subgrid errors (because the electro-motive force in equation (\ref{eq: induction}) is nonlinear), which dominate over measurement errors and must be accounted for \citep{pais2008quasi,baerenzung2016flow}. 
One could add model errors upon measurement errors with no other modification to the algorithms.
However, errors mentionned above are correlated in time \citep{gillet2015planetary}, the reason why they have been considered by the means of stochastic differential equations by \cite{barrois2017contributions}. 
The authors show with an augmented state OI filter how mandatory it is to account for model uncertainties to correctly recover transient surface core motions. 
Their use of a stress-free geodynamo spatial prior \citep{aubert2013bottom} put to the fore a signature of magnetic diffusion on short periods (as it is enslaved to rapid flow changes). 
\cite{baerenzung2017modeling} follow a similar stochastic avenue using instead an EnKF, involving some 10,000 ensemble members to obtain well-conditionned ${\sf P}^f$. 
They emphasize two different time-scales for the dynamics: a millenial planetary gyre superimposed with more rapid and spatially localized features.
These two geophysical applications of data assimilation tools illustrate how mixing constraints from dynamical models and observations allow to extract information about the core physics. 

\subsection{Dynamically Constrained Geomagnetic Field Models}
\label{sec: assim applications}

Early assimilation strategies have been introduced with operationnal perspectives (production of field models, SV predictions). 
\cite{sanchez2016modelling} proposed series of snapshot archeomagnetic field models under a dynamo norm, showing that some information may be retrieved up to spherical harmonic degrees $n=5$, depending on the considered epoch. 
Applications to space weather were anticipated by \cite{aubert2015geomagnetic} who generated decadal forecast of the South-Atlantic anomaly with geodynamo equations, starting from an initial regression of MF and SV data under a dynamo norm.
\cite{beggan2010forecasting} tested the improvement of SV predictions with piece-wise constant flows advecting the core surface field within an EnKF framework.

The production of 5 yrs SV predictions involving assimilation tools has been proposed in the context of the International Geomagnetic Reference Field: \cite{kuang2010prediction} plugged geodynamo equations into univariate OI schemes; \cite{fournier2015candidate} performed multivariate snapshot inferrence implying geodynamo cross-covariances; \cite{gillet2015stochastic} integrated a reduced stochastic QG model into an augmented state Kalman Filter. 
Ultimately, producing unbiased estimates will require mixing dynamos equations, in the bad parameter regime or involving parameterized subgrid processes, together with (stochastic) parameterization of unresolved quantities.  

\section[]{Concluding Remarks}
\label{sec: conclusion}

To close this chapter, I remind some of the main difficulties that we must face in order to take a step forward. 
On long periods, on the top of new records, magnetic field modellers need accurate error estimates (for both measurements and unmodelled processes) associated with archeomagnetic and sediment records, to be incorporated with realistic prior information into the inverse problem. 
On shorter periods, the main barrier is related to external signals that hide the information coming from the core, preventing from resolving rapid changes towards short wave-lengths. 
Overtaking this difficulty will possibly require considering dynamical models of the ionosphere, when inverting for satellite and observatory data.  

Numerical models of the core hardly reach the condition $A<O(1)$. 
To significantly lower $A$ one must resort to subgrid-scale parameterization. 
These have to be invented, since no such model is currently suitable for turbulent rotating MHD, where reverse energy cascades make large length-scale features much sensitive to unresoved circulations. 
In front of such complex issues, there is a need for reduced models, for two reasons. 
First, they help better understand the underlying physics, possibly allowing for short-cuts in the expensive modeling from primitive equations. 
Second, as they rely on a smaller number of parameters, they are ideal for integration into data assimilation algorithms. 
These indeed represent heavily under-determined inverse problems: having only access to large length-scale, low frequency $B_r$ at the CMB, only about 200 observations each year are available to constrain the entire core state.

The implementation of geomagnetic assimilation tools requires realistic measures of the pronostic model imperfections. 
If early stochastic models are developed, their integration into three-dimensional models remains untrodden. 
Similarly, localization methods widely used for re-analyzing surface envelopes (ocean and atmosphere) with moderate ensemble sizes \citep{oke2007impacts} will be important to implement, to propagate at best the information through the succession of analyses. 
This nevertheless requires specific thoughts, as in the Earth's core the thick spherical shell geometry, together with a momentum balance dominated by Coriolis and Lorentz forces, make non-local interactions ubiquitous. 

%------------------------------------------------------------------------------------
\subsection*{Acknowledgements}
%------------------------------------------------------------------------------------

I am grateful to Nils Olsen and Chris Finlay for providing the revised observatory monthly series, and to Sabrina Sanchez for her ensemble of AmR field models. 
Discussions with colleagues helped clarify my mind on several points: archeomagnetic field modeling (Erwan Th\'ebault), stratified layers (J\'er\'emie Vidal, Nathanael Schaeffer and Dominique Jault), geodynamo simulations (the last two plus Julien Aubert).  
I thank David C\'ebron for his comments on the manuscript before submission. 
NG is partially supported by the French Centre National d'Etudes Spatiales (CNES) for the study of Earth's core dynamics in the context of the Swarm mission of ESA. 
ISTerre is part of Labex OSUG@2020 (ANR10 LABX56). 

\bibliography{gillet_refs}
\bibliographystyle{abbrvnat}

\end{document}